\documentclass[twocolumn,prb,superscriptaddress,showpacs,preprintnumbers,amsmath,amssymb]{revtex4}
\usepackage{graphicx}
\usepackage{bm}
\usepackage{amsmath,amsfonts,amssymb} 
\usepackage{verbatim}

\usepackage[noglobalbang]{feyn}

\usepackage{tikz}
\usetikzlibrary{arrows,automata,trees,calc,3d,shapes.multipart,arrows}
\usetikzlibrary{calc,patterns}
\usepackage{natbib}
\begin{document}

\title{Generalized inclusion of short range ordering effects\\
in the coherent potential approximation.}

\author{Alberto Marmodoro}
\affiliation{University of Warwick, Department of Physics, Coventry, CV4 7AL, United Kingdom} 
\affiliation{Max-Planck-Institut f\"ur Mikrostrukturphysik, Weinberg 2, D-06120 Halle, Germany}
\email{amarmodoro@mpi-halle.de} 
\author{Arthur Ernst}
\affiliation{Max-Planck-Institut f\"ur Mikrostrukturphysik, Weinberg 2, D-06120 Halle, Germany}
\email{aernst@mpi-halle.de}
\author{Julie B. Staunton}
\affiliation{University of Warwick, Department of Physics, Coventry, CV4 7AL, United Kingdom} 
\email{j.b.staunton@warwick.ac.uk}
\date{\today}
\begin{abstract}
The coherent potential approximation has historically allowed the efficient study of disorder effects over a variety of solid state systems. Its original formulation is however limited to a single-site or uncorrelated model of local substitutions. This neglects the effects of correlation and short range ordering, often found in realistic materials. Recent theoretical work has shown how to systematically address such shortcomings, for simple materials with only one element per unit cell. We briefly review the basic ideas of these developments within the framework of multiple scattering theory, and suggest their generalization to materials with complex lattices and possibly different types of disorder. We illustrate this extension with an example of local environment effects in the exotic Hapkeite $Fe-Fe_{37.5\%}Si_{62.5\%}$ compound.
\end{abstract}

\pacs{71.23.-k, 71.15.Mb, 71.15.Ap, 71.15.Dx, 71.20.Be, 71.20.Gj, 71.20.Lp, 75.50.Bb, 75.30.Hx, 75.20.Hr}

\maketitle

\section{Introduction}
The first-principles treatment of solid state systems in the presence of disorder poses an extra degree of technical and numerical difficulties, due to loss of crystalline periodicity.~\cite{Gyorffy1991a} Some deviations from such ideal scenario are however to some extent always present: real samples always contain various kinds of defects, which may have a significant influence on the fine details of their electronic structure.

Moreover, the general framework of a substitutional model of disorder, where occupation of any lattice site by alternative species is only determined in probabilistic terms, applies to a variety of other physical problems of great practical interest. In addition to the study of metallic alloys,~\cite{Kirkpatrick1970,Bansil1978,Giuliano1978,Temmerman1978}, this includes among various examples doped semiconductors, but also random spin systems~\cite{Matsubara1973} such as magnetic materials above a critical ordering temperature~\cite{Staunton1984} and further scenarios.

One efficient technique to tackle these problems is provided by the coherent potential approximation~(CPA)~\cite{Soven1967} in its multiple scattering Korringa, Kohn and Rostoker (KKR) implementation.~\cite{Korringa1947,Kohn1954}\cite{Gyorffy1978,Temmerman1987,Gyorffy1991a} 
The theory operates by examining the actual physical system in terms of a computationally more amendable effective medium,
which is determined self-consistently for each energy so as to best
incorporate average disorder effects. It has been shown to satisfy a
strict set of theoretical requirements, related to convergence and
analyticity of a proper solution;~\cite{Mills1978,Mills1983} to recover exact results in all the appropriate limits;~\cite{Velicky1968} and to provide qualitative, systematic improvements over cruder models of disorder~\cite{Levin1970} at a reasonable increase in computational demands.

On the other hand, reliance on concentration alone to model the average presence of different chemical or magnetic species cannot account for possible higher order aspects of their actual distribution, as found in a real material.~\cite{Capek1971} Local atomic substitutions in fact typically extend their effect up to an extended length scale, which may or may not correspond to that of crystalline periodicity, or other forms of long-range order~(LRO) in a sample.

Attempts to improve the original single-site CPA formulation to also describe such short-range order~(SRO) or intermediate disorder scenarios have prompted a large body of theoretical efforts.~\cite{Rowlands2009} On the one side can be placed developments to include the case of complex unit cell materials, where some forms of disorder coexist with undisturbed periodicity on the other sublattices.~\cite{Pindor1983} From a complementary point of view, there have been also attempts to explicitly re-insert local higher order corrections in the original theory for simple systems, on top of the previous average -only results.~\cite{Mills1983}

Along the latter line of research, the non-local coherent potential approximation (NLCPA) has been recently developed based on insights provided by the Dynamical Cluster Approximation (DCA).~\cite{Jarrell2001} The method has shown its capability to describe SRO effects both within model Hamiltonian studies~\cite{Rowlands2003,Rowlands2004,Rowlands2006,Rowlands2008,Rowlands2009} and when reimplemented within a completely self-consistent DFT-KKR framework for realistic systems.~\cite{Rowlands2006a,Tulip2006,Koedderitzsch2007}

This technique inherits however from its derivation a practical restriction to simple, one atom per unit cell lattices in \textit{Strukturberichte} $A_h$, $A2$ or  $A1$ structures. In this paper we reexamine the foundations of such approach, and proceed to generalise it to also include the case of complex unit cell, multiple sublattices scenarios, in arbitrary geometries. Following previous discussion for a model Hamiltonian application,~\cite{Marmodoro2011} this work considers in particular the problem from the prospective of multiple scattering calculations for actual materials. 
Many potential examples spring to mind including high $T_c$ superconductors, multiferroics, candidate compounds for hydrogen storage etc. To illustrate some of the effects which the new method can describe, we presently apply it however to the specific case of the Hapkeite iron-silicon mineral~\cite{Khalaff1974,Anand2003,Hiltl2011}. This alloy is naturally formed in lunar soil by impact-induced, vapor phase deposition, and
turns out to be a particularly illustrative example for current discussion, thanks to its particular lattice simplicity. This follows a $Pm3m$ symmetry, where one of the two distinct crystallographic positions is periodically occupied by $Fe$ atoms, while the other can randomly host either iron or silicon. This kind of intermixing highlights the sensitivity of the transition metal not only to the overall concentration of the $Si$ atoms, but also to specific local environment effects.

The paper is organized as follows. The essential aspects of the original CPA method are briefly reviewed, and its single-site limitations are tracked down in particular to the step where the conditional averages over a variety of local realizations of lattice occupations are evaluated.  Our analysis points out how higher order corrections, already quantified by Mills~\textit{et al.} for model Hamiltonians in the travelling cluster approximation~(TCA),~\cite{Mills1983} are missed out in the original theory. We recap from a similar angle the alternative strategy to reinclude these adopted by the NLCPA (section \ref{sec:review}); and the independent generalisation of the single-site CPA, proposed instead by Pindor~\textit{et al.}~\cite{Pindor1983} to describe complex unit cell cases (MS-CPA, section \ref{sec:MS-CPA}). 
Finally, we present our unified solution for the treatment of LRO and SRO scenarios in complex lattice materials as a formal merger of the MS-CPA and NLCPA (section \ref{sec:extension}).

This multi-sublattice, non-local generalization of the theory (MS-NLCPA) is first put to the test by artificially examining the $A2$ ($A1$) disordered $Cu_{1-c}Zn_c$ metallic alloy as an example of a multi-sublattice $B2$ ($L1_2$) material, and showcasing the full equivalence of our MS-NLCPA results to those of the
original NLCPA~\cite{Rowlands2006} (section~\ref{sec:validation}).


We then demonstrate the new functionality of the method with an application to the Hapkeite $Fe-Fe_{c}Si_{1-c}$ compound. We explore here in particular the connection between electronic and magnetic properties of the iron atoms as a function of their average local environment, under a fixed experimental concentration $c=37.5\%$.

The main aspects of these developments are finally summarized in section \ref{sec:conclusions}.

\section{Effective medium studies of disordered systems}
\label{sec:review}
We briefly review the basic ideas of the theory by referring to a generic binary alloy in the form: $A_{1-c}B_c$. 
Our interest lies with observables that involve a statistically significant number of randomly occupied lattice sites,
over which averaged properties are assessed. Typical experimental techniques include those of photoemission and
nuclear magnetic resonance spectroscopies; electrical and optical transport; estimates of magnetic and compositional transition temperatures; and so on.

In all these cases, it is a great many configurations of atomic species $A$ and $B$ within different local arrangements that give rise to the aggregated, observable result. 

We approach the computational challenge posed by these scenarios by replacing the original problem with an effective medium description. This represents a further step of abstraction in the context already established by density functional theory, in which one-electron -like Kohn-Sham orbitals are evaluated starting from an approximation of the actual many-body potential generated by all charges, in the specific arrangement of various atomic sites on a lattice.

In a multiple scattering solution, these ``scatterers'' are decoupled in the effect produced by local interactions, and the global spatial distribution. Such separation is used to efficiently determine the electronic propagator $\mathcal{G}(\bm k,\epsilon)$, and from it the electronic charge density. This provides the necessary input for a further iteration of the Kohn-Sham scheme, until convergence is achieved and additional observables may be determined~\cite{Johnson1986}.

When some form of disorder disrupts periodicity of occupation over the various lattice sites, the CPA applies to the above general scheme an additional demand that the system Green function should reproduce the average properties of the actual material, evaluated over a large enough portion of the bulk. We briefly review the basics of this technique to also highlight one of its specific limitations, before examining some suggested amendments.

\subsection{The coherent potential approximation and its single-site nature}
In the original derivation by Faulkner and Stocks,~\cite{Faulkner1980} an approximation for the ensemble averaged propagator $\overline{\mathcal{G}}(\bm k,\epsilon)$ is offered in terms of an effective medium scattering path operator $\hat{\overline{\tau}}(\epsilon)$. This mathematical object contains all specific properties for the system, and is efficiently and conveniently represented in a spherical harmonics and lattice site basis, denoted here in its matrix form over the $L=(\ell,m)$ and site $i$ indices as $\underline{\underline{\overline{\tau}}}(\epsilon)$.

The CPA provides a scheme to obtain such quantity from the condition of no extra scattering coming on the average from any embedded impurities.
%
In practice this constraint can be enforced starting for each energy $\epsilon$ from the T-matrix approximation (ATA) Ansatz~\cite{Gonis1992}, which posits for the $t$-matrix associated with every site the average form:
\begin{equation}
  \label{eq: CPA ATA}
\alpha=A \textrm{ or } B
\Rightarrow
  \underline{\overline{t}}(\epsilon) = 
  (1-c) \underline{t}_A (\epsilon) + c \underline{t}_B (\epsilon) 
\end{equation}
If we assume that the CPA effective medium should be on average the same on each lattice position $n$ (as often indicated in the literature through the convention $n=0$, here and in the following), we can adopt a Brillouin zone integral representation and restrict ourselves to the finite site-diagonal part of $\underline{\underline{\overline{\tau}}}(\epsilon)$:
\begin{equation}
\label{eq: BZ CPA}
\overline{\underline{\tau}}^{n,n}(\epsilon)
=
\frac{1}{\Omega}
\int_{\Omega}
d\bm k
\Big(
\underline{\overline{m}}(\epsilon)
-
\underline{G}(\bm k,\epsilon)
\Big)^{-1}
\end{equation}
where $\Omega$ is the Brillouin zone volume, $\underline{\overline{m}}(\epsilon) = \underline{\overline{t}}^{-1}(\epsilon)$, and $\underline{G}(\bm
k,\epsilon)$ represents the structure constants matrix.~\cite{Ham1961}

A self-consistent prescription can then be set up, by also requiring that a concentration weighted sum of the conditional averages over constituents in direct space should also lead to the same result,
\begin{equation}
\label{eq: average CPA}
\overline{\underline{\tau}}^{n,n}(\epsilon)
=
(1-c)
\underline{\tau}^{n,n}_{A}(\epsilon)
+
c
\underline{\tau}^{n,n}_{B}(\epsilon)
=
\sum_{\alpha} 
c_{\alpha}
\underline{\tau}^{n,n}_{\alpha}(\epsilon)
\end{equation}
where $\underline{\tau}^{n,n}_{\alpha}(\epsilon)$ describes an impurity of type $\alpha$, embedded in the CPA medium at site $n$. This quantity can be obtained through application of the corresponding projector:~\cite{Faulkner1980} 
\begin{eqnarray}
\label{eq: SD CPA projector}
\underline{D}_{\alpha}(\epsilon)
=\Big(\underline{1}
+
\overline{\underline{\tau}}^{n,n}(\epsilon)
\Big(
\underline{m}_{\alpha}(\epsilon) - \overline{\underline{m}}(\epsilon) 
\Big)
\Big)^{-1}
\end{eqnarray}
for $\underline{m}_{\alpha}(\epsilon)=\underline{t}^{-1}_{\alpha}(\epsilon)$, so that:
\begin{equation}
\underline{\tau}^{n,n}_{\alpha}(\epsilon)
=
\underline{D}_{\alpha}(\epsilon)
\overline{\underline{\tau}}^{n,n}(\epsilon)
\end{equation}

and iteration between eq.~\ref{eq: BZ CPA} and \ref{eq: average CPA} until both prescriptions converge to the same scattering path operator $\overline{\underline{\tau}}^{n,n}(\epsilon)$ provides the desired CPA description of any average site, within the supposedly infinite bulk.

We now follow Mills \textit{et al.}~\cite{Mills1978} in calling attention to a basic assumption of this approach.
The conditional ensemble average of $N-1$ randomly occupied sites has been so far worked out by simple factorization into $N$ fully uncorrelated substitutions. However when more than one disordered contributions are to be considered together, a proper averaging procedure should also
contain higher order terms. In the generic example of $N$ random
variables $X_1,X_2,\ldots,X_N$, these corrections can be
recursively defined in the form of cumulant averages
(CA):~\cite{Kubo1962,Mills1978,Ziman1979}
\begin{equation}
\label{eq: cumulants expansion}
\begin{array}{lcl}
\langle X_1 \rangle
&=&
\langle X_1 \rangle^C
\\
\langle X_1 X_2 \rangle
&=&
\langle X_1 \rangle \langle X_2 \rangle
 + 
\langle X_1 X_2 \rangle^C
\\
\langle X_1 X_2 X_3 \rangle
&=&
\langle X_1 \rangle \langle X_2 \rangle \langle X_3 \rangle
+ 
\langle X_1 X_2 X_3 \rangle^C
+
\\
&&
+
\langle X_1 X_2 \rangle^C \langle X_3 \rangle^C
+
\langle X_2 X_3 \rangle^C \langle X_1 \rangle^C
+
\\
&& 
+
\langle X_1 X_3 \rangle^C \langle X_2 \rangle^C
\\
\ldots
&=& 
\ldots
\\
\langle X_1 X_2 \ldots X_N \rangle
&=&
\langle X_1 \rangle \langle X_2 \rangle \langle \ldots \rangle \langle X_N \rangle
 + 
\ldots 
\end{array}
\end{equation}
which are rigorously null only in a single site case, and otherwise given by $\langle X_1 X_2 \rangle^C =
\langle X_1 X_2 \rangle
-
\langle X_1 \rangle \langle X_2 \rangle
$, and so forth.

In the present context, these terms reflect the influence that placement of an impurity $\alpha$ on site $n$ exercises on its neighbors.~\cite{Faulkner1980}  In a SRO regime, such an impact decays over a smaller distance than the size of the bulk, and does not correspond to the effective medium periodicity, implicitly assumed in eq.~\ref{eq: BZ CPA}. Significant effects can however arise, which are missed in the cruder single-site factorization of the original CPA theory.

Our approach to overcome such limitation picks up from the non-local extension of the method,~\cite{Rowlands2003} which we now briefly proceed to review.
\subsection{The Non-Local CPA solution}
\label{sec: NLCPA}
The NLCPA operates as well within the basis of a factorized evaluation of the effective medium, combining the contributions from different
substitutional species $\alpha$. 

In this formalism, however, these are now considered over clusters of $N_c \geq 1$ sites, occupied by sets $\gamma = \lbrace \alpha_1, \alpha_2, \ldots,
\alpha_{N_{c}} \rbrace$ of possibly correlated impurities. A multi-site probability distribution $P(\gamma)$ generalizes then the role of the single-site concentration $c_{\alpha}$, and can account for different forms of SRO over a length scale fixed by the size of the cluster.
When $N_{c} = 1$, the method reverts to the CPA treatment, from which it
also inherits all the analytical features of a Herglotz
solution.
It becomes however exact in the limit of cluster size extending to the whole bulk, $N_{c} = N \to \infty$.
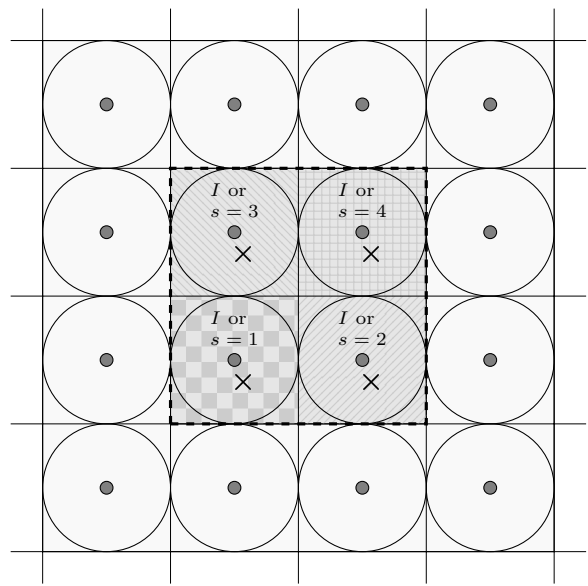
\begin{figure}[htb]
 \centering
\begin{tikzpicture}[scale=0.85]
\draw[step=2cm,fill=gray!5!white] (0,0) rectangle (8,8);
\draw[step=2cm,fill=gray!20!white] (2,2) rectangle (6,6);
\pattern [pattern=checkerboard,pattern color=black!20]
    (2,2) rectangle (4,4);
\pattern [pattern=grid,pattern color=black!20]
    (4,4) rectangle (6,6);
\pattern [pattern=north west lines,pattern color=black!20]
    (2,4) rectangle (4,6);
\pattern [pattern=north east lines,pattern color=black!20]
    (4,2) rectangle (6,4);
\draw[step=2cm,very thin] (-.5,-.5) grid (8.5,8.5);
\def\Rmin{.1};
\def\Rmax{1};
\foreach \x in {1,3,...,8}
{
  \foreach \y in {1,3,...,8}
  {
    \draw(\x,\y) circle (\Rmax);
    \draw[fill=gray] (\x,\y) circle (\Rmin);
  }
}
\draw[step=2cm,very thick,dashed] (2,2) rectangle (6,6);
\foreach \x [evaluate = \x as \xpp]  in {3,5}
{
  \foreach \y [evaluate = \y as \ypp using int(\x/2+(\y/2-1)*2-1)] in {3,5}
  {
    \draw[below right] (\x,\y) node(){$
\Diagram{
x
}$};
    \draw[above] (\x,\y) node(){
\begin{scriptsize}
$\begin{array}{l}
I \textrm{ or } \\
s= \ypp
\end{array}$
\end{scriptsize}
    };
  };
};
\end{tikzpicture}
\caption{Schematic depiction of a multi-site cavity with $N_{c}=4$ elements (darker gray). The cavity is evaluated at an arbitrary position from the effective medium (lighter gray). Various substitutional impurities are placed in different, possibly correlated arrangements $\gamma$, according to a multi-site probability distribution $P(\gamma)$. The different labelling of distinct site occupation pertains to a NLCPA ($\alpha_I$ index, $N_{sub}=1$, $N_c=4$) or MS-NLCPA ($\alpha_{I,s}$ index, $N_{sub}=4$, $N_c=1$) descriptions, which can be equivalently deployed and lead to the same results in \textit{ad hoc} degenerate test cases (see text).}
\label{fig: generalized cavity}
\end{figure}

Technically, this is accomplished through a modified version of eq.~\ref{eq: BZ CPA} and \ref{eq: average CPA}, based on an enlarged but finite matrix structure for the scattering path operator $\underline{\underline{\overline{\tau}}}(\epsilon)$. Here a second underline denotes now a blockwise extension of the original CPA matrix into an extra pair of indices, labeling different $I,J=1,\ldots,N_{c}$ cluster elements in the bulk (Fig.~\ref{fig: generalized cavity}). The corresponding impurity projectors become:

\begin{eqnarray}
\label{eq: NLCPA projectors}
\underline{\underline{D}}_{\gamma}(\epsilon)
=
\left(\underline{\underline{1}}
+
\overline{\underline{\underline{\tau}}}(\epsilon)
\left( \underline{\underline{m}}_{\gamma}(\epsilon) - \underline{\underline{\overline{m}}}(\epsilon) \right)
\right)^{-1}
\end{eqnarray}
for
$\underline{\underline{m}}_{\gamma}(\epsilon)=\underline{\underline{t}}^{-1}_{\gamma}(\epsilon)$ a matrix diagonal in cluster site indices, describing the single site scattering from atoms at sites $I$ within the cluster, according to each particular configuration $\gamma$:
\begin{equation}
\label{eq: D SCF NLCPA usage}
\underline{\underline{\tau}}_{\bm \gamma}(\epsilon)
=
\underline{\underline{D}}_{\gamma}(\epsilon)
\overline{\underline{\underline{\tau}}}(\epsilon)
\end{equation}
Similarly to eq.~\ref{eq: average CPA}, we then assume:
\begin{equation}
\label{eq: NLCPA average}
\underline{\underline{\overline{\tau}}}(\epsilon)
=
\sum_{\gamma} P(\gamma) \underline{\underline{\tau}}_{\gamma}(\epsilon)
\end{equation}
or may equivalently express the contribution of each configuration $\gamma$ in terms of a cavity scattering path operator $\overline{\underline{\underline{\tau}}}^{cav}(\epsilon)$, such that:~\cite{Rowlands2006}
\begin{equation}
\label{eq: NLCPA cavity based projection}
\begin{array}{lcl}
\overline{\underline{\underline{\tau}}}^{cav}(\epsilon)
&=&
\underline{\underline{\overline{m}}}(\epsilon)
 - 
\underline{\underline{\overline{\tau}}}^{-1}(\epsilon)
\\
\underline{\underline{\tau}}_{\gamma}(\epsilon)
&=&
\left(
\underline{\underline{m}}_{\gamma}(\epsilon)
-
\underline{\underline{\overline{\tau}}}^{cav}(\epsilon)
\right)^{-1}
\end{array}
\end{equation}

Care is needed when extending eq.~\ref{eq: BZ CPA} in reciprocal space, to correctly describe a finite cluster still representative for the whole bulk. Jarrell \textit{et al.},\cite{Jarrell2001} and subsequently Rowlands \textit{et al.}, \cite{Rowlands2003} have shown how this can be accomplished through a partitioning of the original Brillouin zone domain of integration respectful of the underlying lattice symmetries. This is now coarse-grained into $N_c$ tiles around a discrete set of cluster momenta $\bm K_n$, which remain defined only up to an arbitrary phase factor, further discussed below~\cite{Rowlands2008}. 

The corresponding scattering path operator is hence obtained in the modified form:
\begin{equation}
\label{eq: NLCPA integral}
\begin{array}{lcl}
\overline{\underline{\tau}}(\bm K_n,\epsilon)
\hspace{-1mm}&=&\hspace{-1mm}
\frac{N_c}{\Omega}
\int_{\Omega_{\bm K_n}}
\hspace{-1mm} d\bm k 
\Big(
\underbrace{
\overline{\underline{t}}^{-1}(\epsilon)
-
\underline{\delta G}(\bm K_n, \epsilon)
}
-
\underline{G}(\bm k,\epsilon)
\Big)^{-1}
\hspace{-.5cm}=\\
\hspace{-1mm}&=&\hspace{-1mm}
\frac{1}{\Omega_{\bm K_n}}
\int_{\Omega_{\bm K_n}}
\hspace{-1mm} d\bm k 
\Big(
\overline{\underline{m}}(\bm K_n,\epsilon)
-
\underline{G}(\bm k,\epsilon)
\Big)^{-1}
\end{array}
\end{equation}
where $\Omega_{\bm K_n}$ stands for the integration volume of each of the $N_c$ tiles, defined so as to preserve on-average translational invariance.~\cite{Jarrell2001} SRO corrections $\underline{\delta G}(\bm K_n, \epsilon)$ for each cluster momenta are made to appear, and can be collected into the now $\bm K_n$ -dependent $\overline{\underline{m}}(\bm K_n,\epsilon)$ 
construct of eq.~\ref{eq: NLCPA integral}, together with the original $\overline{\underline{t}}^{-1}(\epsilon)$.

In practice, the algorithm begins again from a blockwise expression of the original ATA Ansatz of eq. \ref{eq: CPA ATA}, initially without these extra terms. It iterates between the results of eq.~\ref{eq: NLCPA average} and eq.~\ref{eq: NLCPA integral} through the coupled set of lattice Fourier transforms:~\cite{Ziman1979}
\begin{equation}
\begin{array}{lcl}
\label{eq: lattice Fourier transforms}
\overline{\underline{\tau}}(\bm K_n, \epsilon)
&=&
\sum_{J}
\overline{\underline{\tau}}^{IJ}(\epsilon) e^{-i \bm K_n \cdot (\bm R_I - \bm R_j)}
\\
\overline{\underline{\tau}}^{IJ}(\epsilon)
&=&
\frac{1}{N_c}
\sum_{\bm K_n}
\overline{\underline{\tau}}(\bm K_n,\epsilon) e^{+i \bm K_n \cdot (\bm R_I - \bm R_j)}
\end{array}
\end{equation}
until satisfactory convergence to the same effective medium description is achieved. Here $\bm R_I$ denotes the position of a site within the embedded cluster, and the Brillouin zone tile centers $\bm K_n$ are related to these vectors such that $ \frac{1}{N_c}\sum_n e^{i \bm K_n \cdot (\bm R_I - \bm R_j)}= \delta_{I,J}$.

The tiling procedure places however some constraints on the allowed sets of such $N_c$ calculation parameters.~\cite{Jarrell2001} To this date the technique has been developed only for simple lattices with just one element per unit cell, in the case of \textit{Strukturberichte} $A_h$, $A2$, $A1$
geometries.~\cite{Rowlands2003} 
Furthermore, this coarse-graining of the Brillouin zone leads to discontinuities in the $\bm k$ -dependent integrand of Eq.\ref{eq: lattice Fourier transforms}, whenever tile boundaries are crossed.~\cite{Tulip2006,Rowlands2008} We note here however a recent suggestion to overcome such limitation through an additional averaging step over various phase choices, to reobtain a smooth $\bm k$ -dependence~\cite{Rowlands2009}.

Before proceeding to develop the multi-atom per unit cell generalisation of this KKR-NLCPA approach, we shall now briefly recall for clarity also the complementary starting point of the single-site CPA applied to complex lattices. 

\subsection{The multi-sublattice CPA for complex lattices in arbitrary geometries}
\label{sec:MS-CPA}
We consider here the extension of the single-site CPA proposed by Pindor \textit{et al.} for multiple sublattices compounds (MS-CPA).~\cite{Pindor1983} This theory is designed to cover cases of ``periodic disorder'', where a material with $s=1,2, \ldots, N_{sub}$ crystallographic positions in the unit cell may randomly host more than one atomic species per site. A prototype $N_{sub} = 2$ example can be given by the $\alpha_1=A$ or $B$, $\alpha_2=C$ or $D$ abstract compound, in the general formula: $(A_{1-c_1},B_{c_1})-(C_{1-c_2},D_{c_2})$.

In this case, the reciprocal space prescription for the site -diagonal scattering path operator of eq.~\ref{eq: BZ CPA} becomes:~\cite{Pindor1983}
\begin{equation}
\label{eq: MSCPA BZ integration}
\begin{array}{lcl}
\underline{\overline{\tau}}_{s,s'}(\epsilon)
&=&

\frac{1}{\Omega}
\int
d\bm k
\Big(
 \underline{\underline{\overline{t}}}^{-1}(\epsilon)
-
\underline{\underline{G}}(\bm k, \epsilon)
\Big)_{s,s'}^{-1}
\delta_{ss'}
=\\
&=&

\frac{1}{\Omega}
\int 
d\bm k
\Big(
\underline{\underline{\overline{m}}}(\epsilon)
-
\underline{\underline{G}}(\bm k, \epsilon)
\Big)_{s,s'}^{-1}
\delta_{ss'}
=
\underline{\overline{\tau}}_{s}(\epsilon)
\end{array}
\end{equation}
where $\underline{\underline{\overline{t}}}^{-1}(\epsilon)$ is the block-diagonal matrix with elements $\underline{\overline{t}}_s \delta_{s,s'}$ in sub-lattice space, and now $\underline{\underline{G}}(\bm k,\epsilon)$ 
describes the structure constants for the free electron propagation  in the complex lattice.
The self-consistent determination of the effective medium starts once again from 
the ATA Ansatz of eq.~\ref{eq: CPA ATA}, modified to have: 
$\overline{\underline{t}}_{1}(\epsilon) = (1-c_1) \underline{t}_A(\epsilon) + c_1 \underline{t}_B(\epsilon) $, and $\overline{\underline{t}}_{2}(\epsilon) = (1-c_2) \underline{t}_C(\epsilon) + c_2 \underline{t}_D(\epsilon)$, in this generic example of a binary alloy with two sublattices.

This prescription is complemented by considering in direct space:~\cite{Pindor1983}
\begin{eqnarray}
\label{eq: MSCPA average}
\underline{\overline{\tau}}_{1}(\epsilon) 
= (1-c_1) \underline{\tau}^{1}_A(\epsilon)  + c_1 \underline{\tau}^{1}_B(\epsilon) 
\\
\underline{\overline{\tau}}_{2}(\epsilon) 
= (1-c_2) \underline{\tau}^{2}_C(\epsilon)  + c_2 \underline{\tau}^{2}_D(\epsilon) 
\end{eqnarray}
with contributions from the different impurities that may be found on each sublattice 
now given by the projectors:
\begin{equation}
\label{eq: MS-CPA projectors}
\underline{D}_{\alpha_s}(\epsilon)
=
\Big(\underline{1}
+
\underline{\overline{\tau}}^{s}(\epsilon)
\Big( 
\underline{m}_{\alpha_s}(\epsilon) - \underline{\overline{m}}_s(\epsilon) 
\Big)
\Big)^{-1}
\end{equation}
so that in general:
\begin{equation}
\underline{\tau}^{s}_{\alpha_s}(\epsilon)
=
\underline{D}_{\alpha_s}(\epsilon)
\underline{\overline{\tau}}^{s}(\epsilon)
\end{equation}

We note that the formulation remains sub-lattice block-diagonal at all steps, despite the complete $N_{sub} \times N_{sub}$ block matrix inversion appearing in eq.~\ref{eq: MSCPA BZ integration}. 
When compared with the NLCPA of eq.~\ref{eq: NLCPA average}-\ref{eq: NLCPA integral}, it can be seen to describe at most LRO or periodic disorder cases, but no local effects from SRO including those linking the occupancies of the different sub-lattices. As expected, this development also reverts back naturally to the single-site CPA for $N_{sub}=1$. It shares with the original formulation also lack of particular restrictions to specific lattice geometries, another element that we wish to retain in our generalisation proposal.

\section{Merging the two treatments}
\label{sec:extension}
Our strategy for combining respective benefits from the two theories of sections \ref{sec: NLCPA} and \ref{sec:MS-CPA} will be again based on a self-consistent procedure to track multiple length scales in an extended cavity, as depicted for the NLCPA in figure~\ref{fig: generalized cavity}. We still resort in particular to considering multi-site substitutions governed by a richer probability distribution $P(\gamma)$, suitable to describe different forms of short and long range ordering. This general scheme is derived from the NLCPA, and indeed the following analysis will be based on the formal demand that the new generalized theory should recover the results from such starting point, in the appropriate limit.

We intend to also match this extended model of disorder with assumptions similar to those underneath eq.~\ref{eq: MSCPA BZ integration}, where now all terms are carefully preserved and no off-diagonal contributions in sublattice space get discarded.
The resulting richer description of the effective medium contains additional non-diagonal t-matrix terms $\overline{\underline{t}}_{s,s'}(\epsilon)$, similar to the $\overline{\underline{t}}_{IJ}(\epsilon)$ contributions in the
original NL-CPA.~\cite{Rowlands2003} When both developments are combined, the scattering path operator is also extended, to ultimately obtain a matrix $\overline{\underline{\underline{\underline{\tau}}}}(\epsilon)$ now labeled in general by angular momentum, tile and sublattice indices. 

This allows the formalism to track the effects of possibly correlated substitutions over $N_{cutoff} = N_c \times N_{sub}$ elements' length scales. 
Special care must however be given to the important role now played by the off-diagonal blocks of the integrand from eq.~\ref{eq: MSCPA BZ integration}.

We consider here for simplicity a $N_c=1$ example. The general expression for the extended scattering path operator $\overline{\underline{\tau}}_{s s'}(\epsilon)$ can be written as:
\begin{equation}
\label{eq: MS-CPA integration modified}
\begin{array}{lcl}
\overline{\underline{\tau}}_{s s'}(\epsilon)
&=&
\frac{1}{\Omega}
\int_{\Omega} d\bm k
\left(
\overline{\underline{\underline{m}}}(\epsilon)
-
\underline{\underline{\tilde{G}}}(\bm k, \epsilon)
\right)^{-1}_{s s'}
\end{array}
\end{equation}
for $\overline{\underline{\underline{m}}}(\epsilon)=\overline{\underline{\underline{t}}}^{-1}(\epsilon)$ at first given by the usual ATA prescription.

The term $\underline{\underline{\tilde{G}}}(\bm k, \epsilon)$ denotes now however modified structure constants matrix blocks $\underline{\tilde{G}}_{s,s'}(\bm k,\epsilon)$, obtained from the original one by considering the extra sublattice -dependent phase modifier:
\begin{equation}
\label{eq: phase correction generic}
\underline{\tilde{G}}_{s s'}(\bm k, \epsilon)
=
\underline{G}_{s s'}(\bm k, \epsilon)
e^{-i \bm k \cdot (\bm r_{s} - \bm r_{s'})}
\end{equation}
which can be postulated by following Banachiewicz's theorem for the blockwise inversion of a square matrix. Here $\bm r_{s}$ is the position vector of site $s$ within a unit cell, and an intuitive interpretation for the action of such correction can be simply illustrated considering the two cases of \textit{Strukturbericht} $A2$ and $B2$ lattices.

In this latter geometry, the off-diagonal structure constants matrices $\underline{G}_{s \neq s'}^{CsCl}(\bm k,\epsilon)$ are related to the corresponding ones for a $BCC$ case according to \cite{Pindor1983}:
\begin{equation}
\label{eq: Pindor relationship}
\underline{G}_{1,2}^{CsCl}(\bm k, \epsilon)
=
\left(
\underline{G}^{BCC}(\bm k, \epsilon)
-
\underline{G}_{1,1}^{CsCl}(\bm k, \epsilon)
\right)
e^{+i \bm k \cdot ( \bm r_{1} - \bm r_{2} )}
\end{equation}
Applying eq.~\ref{eq: phase correction generic} leads hence to obtain, for the non trivial off-diagonal blocks:
\begin{equation}
\begin{array}{lcl}
\underline{\tilde{G}}_{1,2}^{CsCl}(\bm k, \epsilon)
&=&
\underline{G}_{1,2}^{CsCl}(\bm k, \epsilon)
e^{-i \bm k \cdot ( \bm r_{1} - \bm r_{2} )}
=\\
&=&
\underline{G}^{BCC}(\bm k, \epsilon)
-
\underline{G}_{1,1}^{CsCl}(\bm k, \epsilon)
\end{array}
\end{equation}
where $\bm r_1 = (0,0,0)$, $\bm r_2 = (1/2,1/2,1/2)$ are possible basis vectors for the complex unit cell of an $A2$ lattice.

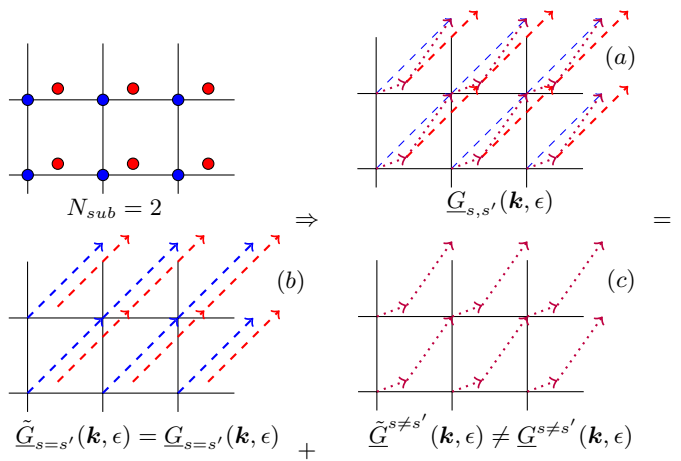
\begin{figure}[htb]
 \centering
\begin{tabular}{ccc}
\hspace{-.75cm}
\begin{tikzpicture}[scale=.5]
\draw[step=2cm] (-.5,-.5) grid (5.5,3.5);
\foreach \x in {1,3,...,5}
{
  \foreach \y in {1,3,...,4}
  {
    \draw[fill=blue] (\x-1,\y-1) circle (.15cm);
    \draw[fill=red] (\x-.2,\y-.7) circle (.15cm);
  }
}
\draw[black] (2.3,-.4) node [below] {$N_{sub}=2$};
\end{tikzpicture}
&$\Rightarrow$&
\hspace{.25cm}
\begin{tikzpicture}[scale=.5]
\draw[black] (6.5,3.5) node [below] {$(a)$};
\draw[black] (3.3,-.35) node [below] {$\underline{G}_{s,s'}(\bm k,\epsilon)$};
\draw[step=2cm] (-.5,-.5) grid (5.5,3.5);
\foreach \x in {1,3,...,5}
{
  \foreach \y in {1,3,...,4}
  {
\draw[->, thin, dashed, blue,line join=round]
     (\x-1,\y-1) -- node [sloped,below] {}
     (\x+1,\y+1);
\draw[->, thick, dashed, red,line join=round]
     (\x-.2,\y-.7) -- node [sloped,below] {}
     (\x+1.7,\y+1.2);
\draw[->, thick, dotted, purple,line join=round]
     (\x-1,\y-1) -- node [sloped,below] {} 
     (\x-.2,\y-.7);
\draw[->, thick, dotted, purple,line join=round]
     (\x-.2,\y-.7) -- node [sloped,below] {} 
     (\x+1,\y+1);
  }
}
\end{tikzpicture}
=\\
\begin{tikzpicture}[scale=.5]
\draw[black] (7,3.5) node [below] {$(b)$};
\draw[black] (3.2,-.5) node [below] {$\underline{\tilde{G}}_{s=s'}(\bm k,\epsilon) = \underline{G}_{s=s'}(\bm k,\epsilon)$};
\draw[step=2cm] (-.5,-.5) grid (5.5,3.5);
\foreach \x in {1,3,...,5}
{
  \foreach \y in {1,3,...,4}
  {
\draw[->, thick, dashed, blue,line join=round]
     (\x-1,\y-1) -- node [sloped,below] {}
     (\x+1,\y+1);
\draw[->, thick, dashed, red,line join=round]
     (\x-.2,\y-.7) -- node [sloped,below] {}
     (\x+1.7,\y+1.2);
  }
}
\end{tikzpicture}
\hspace{-.5cm}
&+&
\begin{tikzpicture}[scale=.5]
\draw[black] (6.5,3.5) node [below] {$(c)$};
\draw[black] (3.3,-.35) node [below] {$\underline{\tilde{G}}^{s\neq s'}(\bm k,\epsilon) \neq \underline{G}^{s\neq s'}(\bm k,\epsilon)$};
\draw[step=2cm] (-.5,-.5) grid (5.5,3.5);
\foreach \x in {1,3,...,5}
{
  \foreach \y in {1,3,...,4}
  {
\draw[->, thick, dotted, purple,line join=round]
     (\x-1,\y-1) -- node [sloped,below] {} 
     (\x-.2,\y-.7);
\draw[->, thick, dotted, purple,line join=round]
     (\x-.2,\y-.7) -- node [sloped,below] {} 
     (\x+1,\y+1);
  }
}
\end{tikzpicture}
\end{tabular}
\caption{Schematic separation into distinct propagation modes for
  purely sublattice diagonal ($s=s'$, $(b)$) and off-diagonal ($s \neq
  s'$, $(c)$) free electron hopping processes, in a $N_{sub}=2$
  sublattices case (blue and red).  Different length scales can be
  tracked consistently with the comparable NLCPA treatment of a
  $N_c=2$ simple lattice scenario, by decomposing the aggregated
  results of panel $(a)$ into the separate contributions of panels
  $(b)$ and $(c)$.  }
\label{fig: phase corrected structure constants}
\end{figure}

Such cancellation of the exponential from eq.~\ref{eq: Pindor relationship} operates in other words by removing the sublattice -diagonal free propagation modes
$\underline{G}_{11}^{CsCl}(\bm k, \epsilon) =
\underline{G}_{22}^{CsCl}(\bm k, \epsilon)$, from the envelope of a generic expression for $\underline{G}_{s, s'}(\bm k,\epsilon)$,
initially designed to account for both diagonal and off-diagonal
hopping processes (Fig.~\ref{fig: phase corrected structure constants}).

For the $s=s'$ terms, there is no exponential argument and the original expression of panel $(a)$ remains unchanged. The suggested modification of eq.~\ref{eq: phase correction generic} however ensures that, when integrating also for the new $s\neq s'$ contributions, such blocks of the scattering path operator keep resolving well distinct length scales effects which may be sensible to sublattice-sublattice couplings. Short range ordering can then be set-up in direct space by straightforward combination of eq.~\ref{eq: NLCPA projectors}-\ref{eq: NLCPA average} and \ref{eq: MSCPA average}-\ref{eq: MS-CPA projectors}, as detailed below.

We find thus in general: 
\begin{equation}
\begin{array}{lcl}
\label{eq: msnlcpa BZ integration with phase}
\underline{\overline{\tau}}_{s,s'}(\bm K_n,\epsilon)
\hspace{-.15cm}&=&\hspace{-.15cm}
\frac{N_c}{\Omega}
\int_{\Omega_{\bm K_n}}
d\bm k
\left(
  \underline{\underline{\overline{t}}}(\bm K_n, \epsilon)
  -
  \underline{\underline{\tilde{G}}}(\epsilon,\bm k)
\right)^{-1}_{s,s'}	
\\
\underline{\overline{\tau}}_{I,s;J,s'}(\epsilon)
\hspace{-.15cm}&=&\hspace{-.15cm}
\frac{1}{N_c}
\sum_{n=1}^{N_c}
\underline{\overline{\tau}}_{s,s'}(\bm K_n,\epsilon)
e^{ + i \bm K_n \cdot (\bm R_I - \bm R_J)}
\end{array}
\end{equation}
as a result of applying, if required, the NLCPA lattice Fourier transforms of eq.~\ref{eq: lattice Fourier transforms} with $\bm R_I, \bm R_J$ and $\bm K_n$ now referring to superlattice vectors.

This larger scattering path operator $\underline{\underline{\underline{\overline{\tau}}}}(\epsilon)$ is again determined self-consistently by solving for contributions from different extended cavity occupations $\gamma = \lbrace \alpha_{1,1}, \ldots, \alpha_{I,s},\ldots,
\alpha_{N_c,N_{sub}} \rbrace$, through a generalization of eq.~\ref{eq: NLCPA projectors}-\ref{eq: MS-CPA projectors}, or the equivalent extension of eq.~\ref{eq: NLCPA cavity based projection}:
\begin{equation}
\begin{array}{lcl}
\label{eq: msnlcpa direct space}
\underline{\underline{\underline{\overline{\tau}}}}^{cav}
(\epsilon)
&=&
\underline{\underline{\underline{\overline{t}}}}^{-1}
(\epsilon)
 - 
\underline{\underline{\underline{\overline{\tau}}}}^{-1}
(\epsilon)
\\
\underline{\underline{\underline{\tau}}}_{\gamma}(\epsilon)
&=&
\left(
\underline{\underline{\underline{t}}}^{-1}_{\gamma}(\epsilon)
-
\underline{\underline{\underline{\overline{\tau}}}}^{cav}(\epsilon)
\right)^{-1}
\end{array}
\end{equation}
so that as in eq.~\ref{eq: NLCPA average}:
\begin{equation}
 \underline{\underline{\underline{{\overline{\tau}}}}}
(\epsilon)
=
\sum_{\gamma} 
P(\bm \gamma)
\underline{\underline{\underline{\tau}}}_{\bm \gamma}
(\epsilon)
\end{equation}
Eqs.~\ref{eq: msnlcpa BZ integration with
  phase}-\ref{eq: NLCPA average} are then iterated
self-consistently, until convergence to the same effective medium
description is reached, and other observables of interest may be computed.

\subsection{Validation tests}
\label{sec:validation}

We propose that a practical approach to validating the above procedure can be conceived by purposedly examining simple one atom per unit cell lattices, taken as particular instances of multi-atom per unit cell materials on a superlattice. We then require that our MS-NLCPA should reproduce the original NLCPA results in this degenerate limit, over all equivalent SRO scenarios.
To this end we consider for instance a typical $Cu_{50\%}Zn_{50\%}$ alloy~\cite{Rowlands2006a,Tulip2006} in either a BCC ($A2$) or FCC ($A1$) phase, and explore different local environment regimes with the two techniques.

The set of fully comparable structures are reported in table \ref{tab: KKR msnlcpa equivalences}. A ``degenerate'' $B2$ setup,
equivalently hosting either $Cu$ or $Zn$ atoms on both sublattices ($N_{sub}=2$, $N_c=1$), can also be examined in a NLCPA study
of a $A2$ unit cell ($N_{sub}=1$, $N_c=2$). Similarly, a $L1_2$ system ($N_{sub}=4$, $N_c=1$) can also be set up to mimic the
equivalent ($N_{sub}=1$, $N_c=4$) $A1$ NLCPA setup. In each case the results should be identical. 

\begin{figure}[htb]
\includegraphics[height=1.75in]{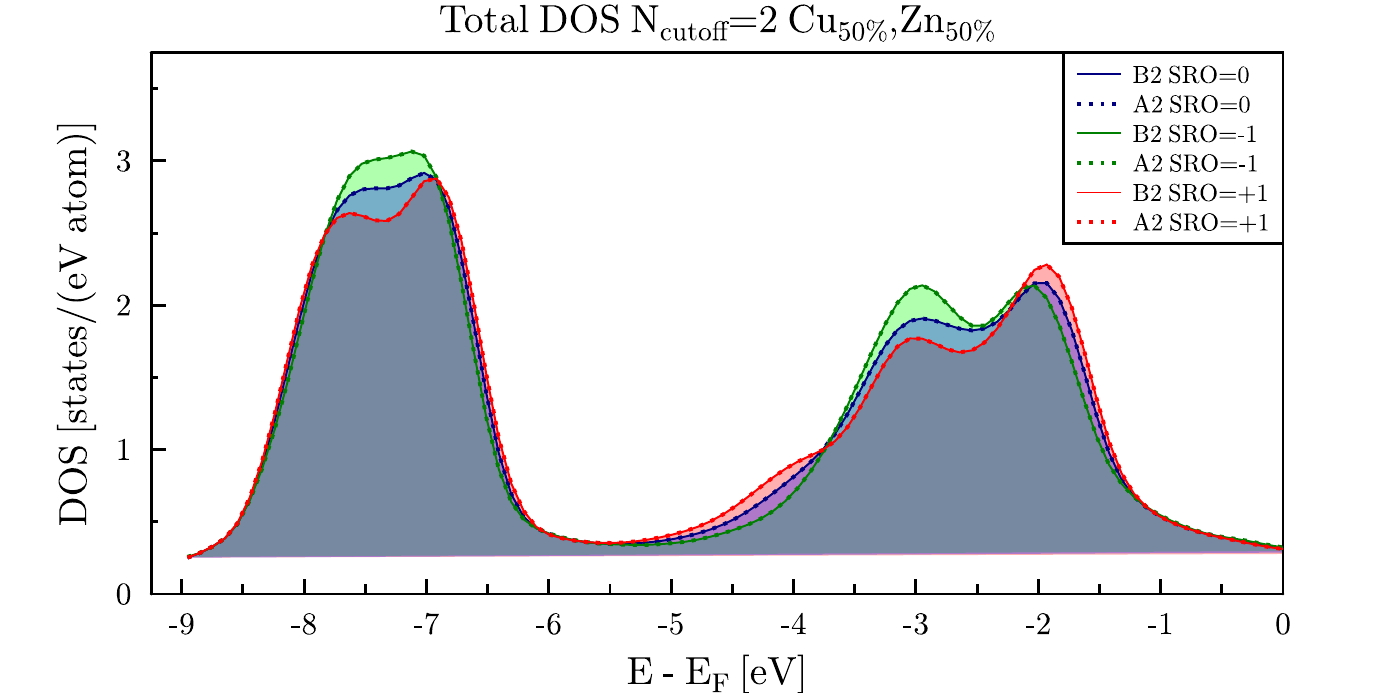}
\caption{The density of states versus energy (color online) for a
  $Cu_{50\%}Zn_{50\%}$ alloy in both a $BCC$ and $CsCl$ structures,
  evaluated for $N_{sub} \times N_c=2$ in the 3 extremal short-range
  ordering regimes of no correlation ($SRO=0$, in blue: $P(\bm
  \gamma_1=\lbrace Cu, Cu \rbrace)=25\%$, $P(\bm \gamma_2=\lbrace Zn,
  Cu \rbrace)=25\%$, $P(\bm \gamma_3=\lbrace Cu, Zn \rbrace)=25\%$ and
  $P(\bm \gamma_4=\lbrace Zn, Zn \rbrace)=25\%$); complete bias towards alike neighbours (clustering, $SRO=+1$, in red: $P(\bm \gamma_1=\lbrace
  Cu, Cu \rbrace)=50\%$ and $P(\bm \gamma_4=\lbrace Zn, Zn
  \rbrace)=50\%$); complete bias towards unlike neighbours (ordering,
  $SRO=-1$, in green: $P(\bm \gamma_2=\lbrace Zn, Cu \rbrace)=50\%$
  and $P(\bm \gamma_3=\lbrace Cu, Zn \rbrace)=50\%$).  The NLCPA results are 
the dotted lines whereas the continuous line shows those of the MS-NLCPA.
 In all $3$ cases, the MS-NLCPA $CsCl$ results
are indistinguishable from the $BCC$ NLCPA ones. }
\label{fig: MSNLCPA budget2 KKR}
\end{figure}

\begin{figure}[htb]
\includegraphics[height=1.75in]{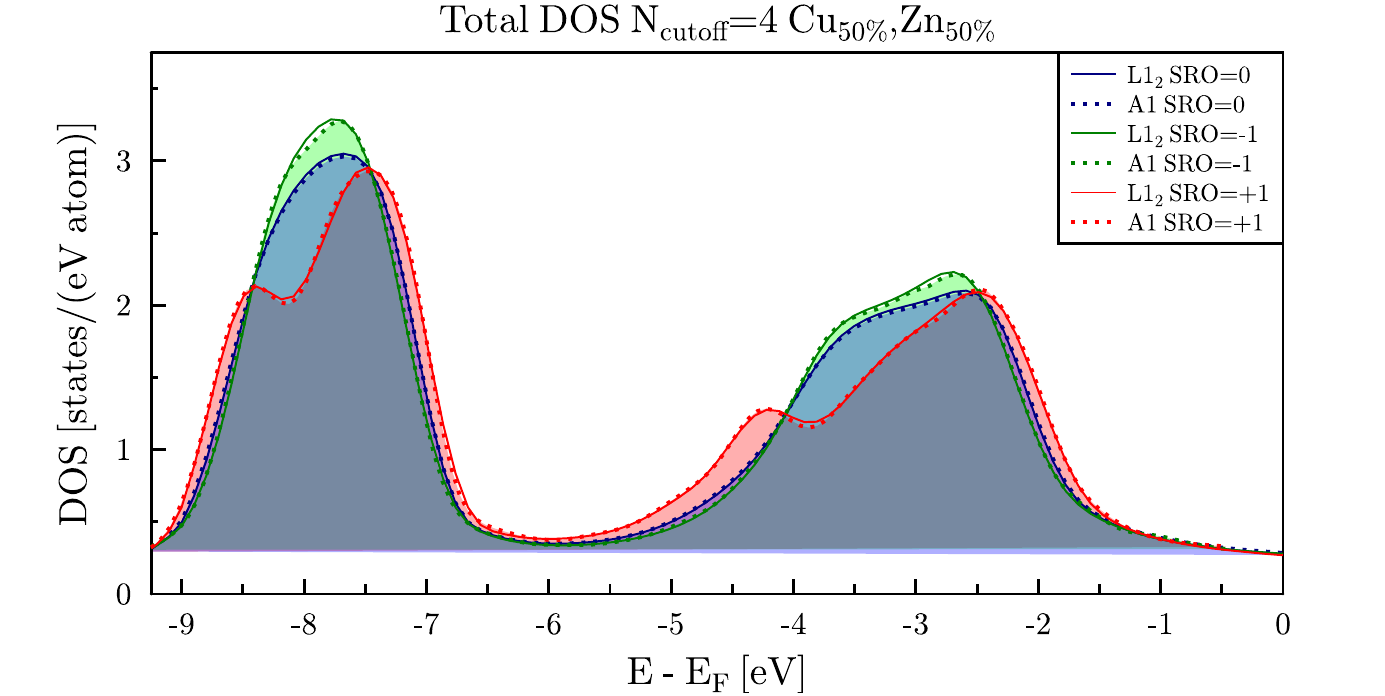}
  \caption{The density of states (color online) for a $Cu_{50\%}Zn_{50\%}$ alloy in both a $FCC$ and $Cu_3Au$ ($L1_2$) structure, evaluated for $N_{sub} \times N_c=4$ in the 3 extremal short-range ordering regimes of no correlation ($SRO=0$, in blue: $P(\bm \gamma_1=\lbrace Cu, Cu, Cu, Cu \rbrace) = P(\bm \gamma_2=\lbrace Cu, Cu, Cu, Zn \rbrace)= \ldots = P(\bm \gamma_{16}=\lbrace Zn, Zn, Zn, Zn \rbrace)=6.25\%$); complete bias towards alike neighbours (clustering, $SRO=+1$, in red: $P(\bm \gamma_1=\lbrace Cu, Cu, Cu, Cu \rbrace)=50\%$ and $P(\bm \gamma_2=\lbrace Zn, Zn, Zn, Zn
    \rbrace)=50\%$); complete bias towards unlike neighbours (ordering, $SRO=-1$, in green: $P(\bm \gamma_1=\lbrace Cu, Zn, Cu, Zn \rbrace)=50$, $P(\bm \gamma_2=\lbrace Zn, Cu, Zn, Cu \rbrace)=50\%$). The NLCPA results are the dotted lines whereas the continuous line shows those of the MS-NLCPA. In all $3$ cases, the MS-NLCPA $Cu_3Au$ results are indistinguishable from the $FCC$ NLCPA ones.}
\label{fig: MSNLCPA budget4 KKR}
\end{figure}

\begin{table}[htp]
 \centering
\begin{tabular}{|c|c|}
 \hline
Computational budget:
 &
Comparable descriptions:
\\
(Nr. of lengthscales)
&
\begin{tabular}{ccc}
Lattice type: & $\quad N_c$ & $N_{sub}$ \\ 
\end{tabular}
\\
\hline
\hline
2
&
\begin{tabular}{ccc}
$BCC$ or $A2$  & $\quad 2$     & $\quad 1$ \\
$CsCl$ or $B2$ & $\quad 1$     & $\quad 2$ \\
\end{tabular}
\\
\hline
4
&
\begin{tabular}{ccc}
\hspace{-.4cm} $FCC$ or $A1$      & $\quad 4$     & $\quad 1$ \\
\hspace{-.4cm} $Cu_3Au$ or $L1_2$ & $\quad 1$     & $\quad 4$ \\
\end{tabular}
\\
\hline
\end{tabular}
\caption[Validation of the MS-NLCPA in a first-principles, KKR framework: comparable calculation setups]{Comparable structural decompositions of the same two physical systems (first and second row). When each sublattice is constrained to host the same elements with the same probabilistic distribution and over matching lattice parameters, the NLCPA and MS-NLCPA are expected to coincide at the level of all effective medium quantities.} 
\label{tab: KKR msnlcpa equivalences}
\end{table}
This equivalence is confirmed for all the relevant effective medium quantities throughout the self-consistent calculation. Following Rowlands \textit{et al.}~\cite{Rowlands2006a}, we show here in particular the density of states for three very different SRO prescriptions for the two physical cases described in columns $1$ and $2$ of table \ref{tab: KKR msnlcpa equivalences} (Fig.~\ref{fig: MSNLCPA budget2 KKR} and \ref{fig: MSNLCPA budget4 KKR}).

In all cases, the $SRO=0$ regime of a fully uncorrelated probability distribution
$P^{SRO=0}(\bm \gamma) = \prod_{I=1}^{N_c}\prod_{s=1}^{N_{sub}}
c(\alpha_{I,s}(\bm \gamma))$ corresponds to a factorized evaluation
of the various elements' concentrations $c$ throughout the cavity, for
$\alpha_{I,s}(\bm \gamma)$ the atomic species found on site $I,s$.  As
already observed in the NLCPA development, this model of disorder
shows in general little difference from the outcome of straightforward single-site CPA calculations.

Additionally however, the extremal cases of full, local clustering ($SRO=+1$):
\begin{equation}
\label{eq: SRO=+1}
P^{SRO=+1}(\bm \gamma)
=
c(\alpha(\bm \gamma)) \textrm{ if } \alpha_{I,s}= \alpha \, \forall \, I,s
\end{equation}
or full local ordering ($SRO=-1$; details in figures captions) may also be successfully compared, as well as any other types of local environments.~\cite{Rowlands2003} The new MS-NLCPA appears to incorporate the NLCPA results as desired.

\subsection{Application to the B2 $Fe-Fe_{37.5\%}Si_{62.5\%}$ Hapkeite mineral}

As more realistic application of the new method, we examine now the example of the Hapkeite phase of the $Fe-Si$ system~\cite{Khalaff1974,Anand2003,Hiltl2011}.

Our choice of such compound is partly based on the desire for a relatively simple yet representative case study to demonstrate the new capabilities of the formalism. The above mineral is characterized by a space group $Pm\overline{3}m$ ($a_{lat}=2.81 \AA$) closely packed structure~\cite{Anand2003}, with a $N_{sub}=2$ complex unit cell hosting $Fe$ on the first sublattice $\bm r_1=(0,0,0)$, and either $Fe$ or $Si$ in the body central position $\bm r_2=(0.5, 0.5, 0.5)$. 

We focus here in particular on the effect of SRO, for a fixed experimental concentration ratio of $Fe:Si=3:5$ on the disordered sublattice, the other sublattice being regularly occupied by iron atoms.
Three scenarios of completely uncorrelated disorder ($SRO=0$), short-range clustering ($SRO=+1$) and local ordering ($SRO=-1$) are again explored, similarly to the validation studies discussed in Fig.~\ref{fig: MSNLCPA budget2 KKR} and \ref{fig: MSNLCPA budget4 KKR}.

We set the maximal correlation length on which such possibilities are considered to the size of a multi-site cavity $N_{cutoff}^{dis}=8$.
This setup can be easily obtained in direct space as the $2\times2\times2$ supercell of the original $B2$ structure. In a fully clustering scenario, the $N_{tot}=2^8=256$ local configurations $\gamma_1 = \lbrace Fe^{s=1},\ldots,Fe^{s=8},\alpha^{s=9},\ldots,\alpha^{s=16} \rbrace$ for $\alpha^s=Fe$ or $Si$, that specify all the possible fillings, are restricted to just $N_{tot}^{SRO=+1}=2$ cases: $\gamma^{SRO=+1}_{1}= \lbrace Fe^{s=1},\ldots,Fe^{s=8},Fe^{s=9},\ldots,Fe^{s=16} \rbrace$ with probability $P(\gamma^{SRO=+1}_1)=0.375$; and $\gamma^{SRO=+1}_{2}=\lbrace Fe^{s=1},\ldots,Fe^{s=8},Si^{s=9},\ldots,Si^{s=16} \rbrace$ with probability $P(\gamma^{SRO=+1}_2)=0.625$. Alternatively, full local ordering can be defined as the restriction to those configurations, where strictly as many $Si$ atoms as mandated by the concentration can occupy the central positions, and all the other ones are hosting iron. The desired elements' ratio is then recovered by giving to each of such $N_{tot}^{SRO=-1}=\left(
\begin{array}{c}
8\\
3
\end{array}\right) = 56
$ arrangements 
a uniform probability distribution: $P(\gamma^{SRO=-1}_1)=\ldots=P(\gamma^{SRO=-1}_{56})=1/N_{tot}^{SRO=-1} \simeq 0.018$. Finally, the case of no restrictions corresponds to the fully uncorrelated, $SRO=0$ regime, which is implicitly assumed in the original MS-CPA.

\begin{figure}[htb]
\centering
\includegraphics[width=\columnwidth]{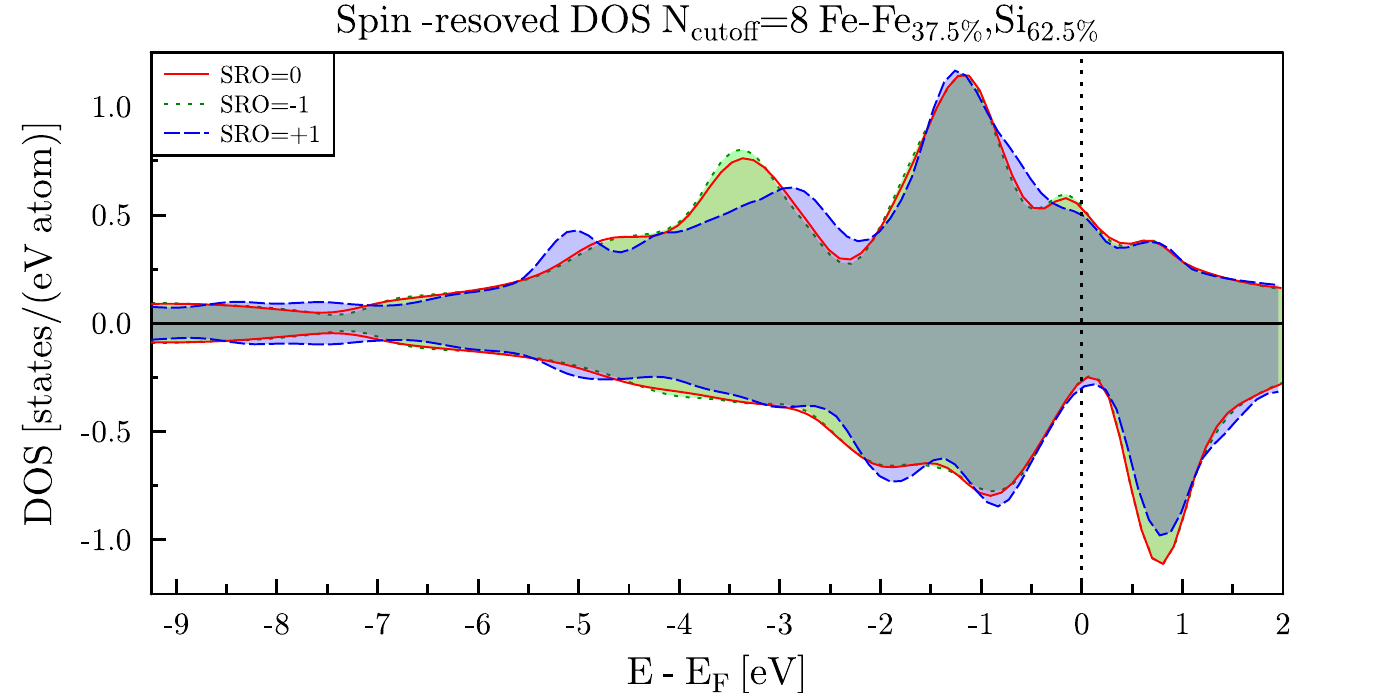}
\caption{Density of states for the Hapkeite mineral $Fe-Fe_{37.5}Si_{62.5}$ in the $B2$ phase, with one sublattice periodically occupied by $Fe$, the second one hosting either $Fe$ or $Si$ under different SRO assumptions (color online). We contrast in particular a fully uncorrelated ($SRO$=0, red) probability distribution of constituents with the case of local ordering ($SRO$=-1, green) or local clustering ($SRO$=+1, blue).}
\label{fig: dos Fe2Si}
\end{figure}

In all these scenarios, a full deployment of the formalism~\cite{Rowlands2006a} in the KKR-CPA scheme generates a new charge distribution associated with each substitutional configuration $\gamma_i$, through iteration until convergence of a standard Kohn-Sham LDA-DFT scheme. This allows to compute element -resolved magnetic moment variations associated with disorder and SRO, as well as other observables such as in this case the spin-resolved DOS. These results are shown in Fig.~\ref{fig: dos Fe2Si}.

\begin{figure}[htb]
\centering
\includegraphics[width=\columnwidth]{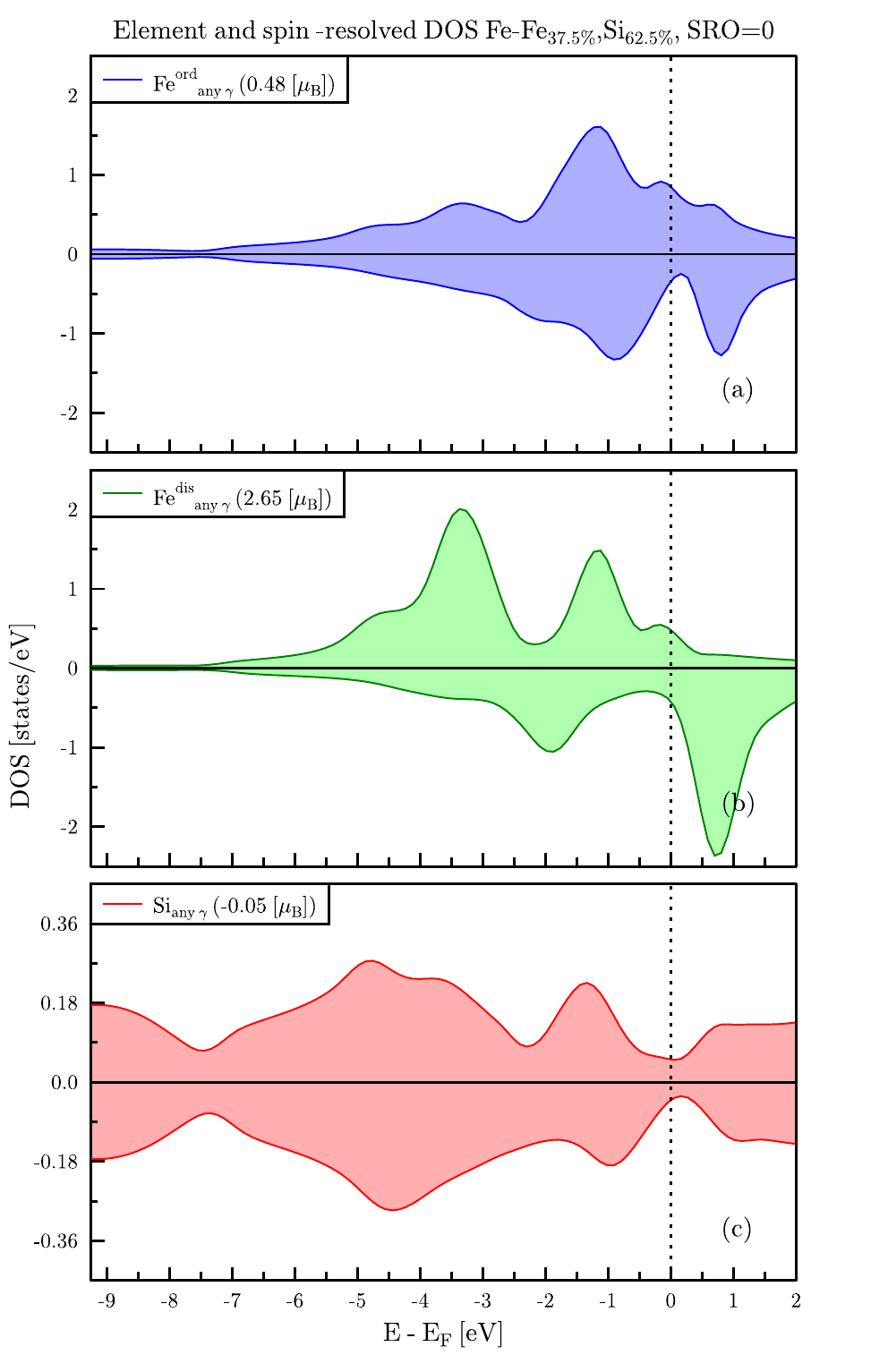}
\caption{(color online) Atomic resolved densities of states and local magnetic moments for the $Fe-Fe_{37.5\%}Si_{62.5\%}$ alloy in the fully uncorrelated case ($SRO=0$): sublattice periodically occupied by $Fe$ (a); sublattice hosting either by $Fe$ (b) or $Si$ (c) atoms.}
\label{fig: dos Fe2SiSRO0}
\end{figure}

\begin{figure}[htb]
\centering
\includegraphics[width=\columnwidth]{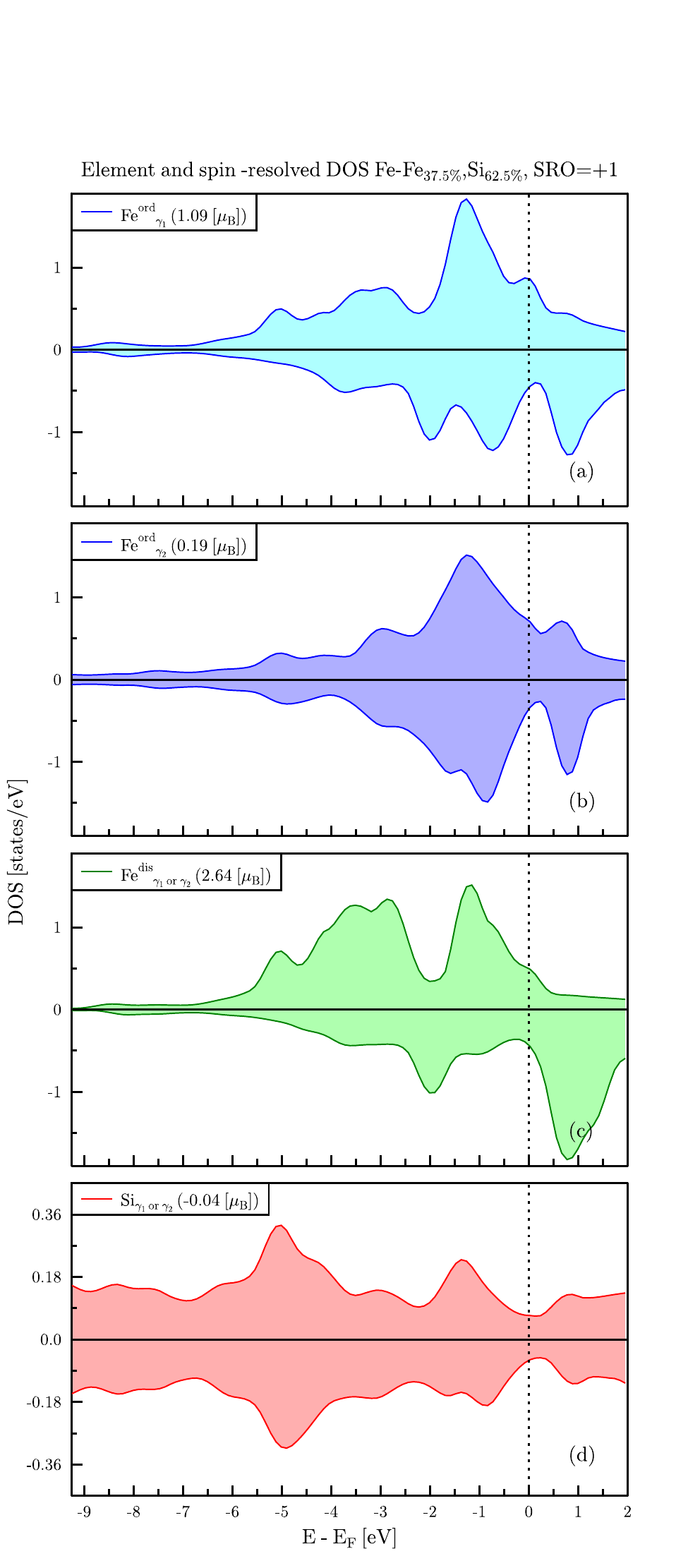}
\caption{(color online) Atomic resolved densities of states and local magnetic moments for the $Fe-Fe_{37.5}Si_{62.5}$ alloy in the case of full clustering ($SRO=+1$, blue): sublattice periodically occupied by $Fe$ in $\gamma^{SRO=+1}_{1}$ (a) and $\gamma^{SRO=+1}_{2}$ (b) configurations; sublattice hosting either by $Fe$ (c) or $Si$ (d) atoms.}
\label{fig: dos Fe2SiSRO1}
\end{figure}

We observe most significant variations for the clustering scenario ($SRO=+1$, Fig.~\ref{fig: dos Fe2SiSRO1}), as opposed to the other two. The most severe constraint on local microscopic compound realizations (from $N_{tot}^{SRO=0}=256$ down to $N_{tot}^{SRO=+1}=2$ alternatives) translates into DOS plots with a more defined outlook, in terms of sharper features across the whole energy spectrum. This can be easily understood noting how the limiting case of complete periodicity or ideal LRO would in fact appear when deploying the method for a $N_{tot}=1$ setup, thereby fully suppressing disorder.
Conversely, the local ordering case $SRO=-1$ appears to be quite sufficient in capturing the essence of electronic structure contributions from the two differently occupied sublattices. 
Results from an uncorrelated model of disorder ($SRO=0$, Fig.~\ref{fig: dos Fe2SiSRO0}) are in fact closely mimicked by evaluating the $56$ contributions of variously permuted $Fe$ and $Si$ elements in the appropriate lattice sites for this locally ordered choice.

The study can lead however to further insight as well into the local environment-resolved impact of distinct constituents. Considering for example the atomic magnetic moment, the fully uncorrelated MS-CPA case provides reference values of $0.48$ $[\mu_B]$ for the iron atoms $Fe_{ord.}^{SRO=0}$ on the completely $Fe$-filled sublattice, whereas those hosted on the other sublattice in competition with silicon atoms, $Fe_{dis.}^{SRO=0}$, have a larger moment of $2.65$ $[\mu_B]$. 


These properties change quite markedly when clustering is instead imposed ($SRO=+1$, Fig.~\ref{fig: dos Fe2SiSRO1}). We note that, while the  magnetic moment of the disordered sites remains substantially unchanged ($2.64$ $[\mu_B]$ for $Fe_{dis.}^{SRO=+1}$), 
the $Fe_{ord.}^{SRO=+1}$ atoms on the other sublattice 
acquire a magnetic moment of $1.09$ $[\mu_B]$ when surrounded again by $Fe$ (configuration $\gamma^{SRO=+1}_{1}$), but only $0.19$ $[\mu_B]$ in a $Si$-only local environment (configuration $\gamma^{SRO=+1}_{2}$). A strong hybridization between electronic levels in the former case leads in fact to the enhancement of local magnetic moment of the periodically repeated $Fe_{ord.}^{SRO=+1}$ atoms, as highlighted in the corresponding spin -resolved DOS shown in Fig.~\ref{fig: dos Fe2Si}. 

Further differences are evident in the $SRO=-1$ scenario. A similar self-consistent calculation attributes to $Fe_{ord.}^{SRO=-1}$ atoms a magnetic moment of $0.58$, $0.52$ or $0.48$ $[\mu_B]$, emerging as clearly distinct values across the $56$ local configurations that are now taken into account.
This can be correlated with the corresponding realization of three specific types of local geometries.
The highest magnetic moment is for instance associated with those $8$ cases where complete iron layers along the [111] direction (and equivalent ones such as $[\overline{1},1,1]$ etc.) are formed within the $2\times2\times2$ supercell. 
The lowest magnetic moment is found instead for local arrangements realizing any of the $6\times4=24$ $(\overline{1},1,0)$ (and permutations) iron layers, which also include a $Si_{Fe}$ defect; and a similar analysis can also be repeated for the intermediate value of $0.52$ $[\mu_B]$. 
For the non-periodic $Fe_{dis.}^{SRO=-1}$ atoms, the magnetic moment appears also close to the $SRO=0$ value, with less significant local environment-dependent variations around an average value of $2.69$ $[\mu_B]$.

\section{Conclusions}
\label{sec:conclusions}
Most materials are affected by some degree of disorder and in many crystalline systems this affects differently a subset of the sublattices. Often properties nominally associated with the chemical composition of one sublattice can be modified by the extent and nature of disorder on another. This effect and more general cases of short range ordering regimes can be overlooked by the original CPA, due to the single-site nature of the theory. In this paper we briefly reexamine further extension to this general method, and suggest a combined approach to efficiently perform first-principles studies on this class of systems.

Previous results for a reference $Cu_{50\%},Zn_{50\%}$ alloy in a $A2$ ($A1$) / $B2$ ($L1_2$) structure could be used to provide strict validation of the new formalism. Furthermore, this could be put to test in its new capabilities, considering here the particularly simple example of a $Fe-Fe_{37.5\%}Si_{62.5\%}$ Hapkeite compound examined from the point of view of basic electronic and magnetic effects induced by different forms of SRO.

\section{Acknowledgment}
Financial support for parts of this work was provided by a University of Warwick, Post-graduate Research Scholarship (WPRS), grants from the Deutscher Akademischer Austausch Dienst (DAAD), and the DFG priority program SPP 1538. 
We wish to express our gratitude to Dr. Sergei Ostanin and Prof. Balasz Gy\"orffy for insightful discussions.

\bibliography{./Bibliography}

\begin{thebibliography}{37}
\expandafter\ifx\csname natexlab\endcsname\relax\def\natexlab#1{#1}\fi
\expandafter\ifx\csname bibnamefont\endcsname\relax
  \def\bibnamefont#1{#1}\fi
\expandafter\ifx\csname bibfnamefont\endcsname\relax
  \def\bibfnamefont#1{#1}\fi
\expandafter\ifx\csname citenamefont\endcsname\relax
  \def\citenamefont#1{#1}\fi
\expandafter\ifx\csname url\endcsname\relax
  \def\url#1{\texttt{#1}}\fi
\expandafter\ifx\csname urlprefix\endcsname\relax\def\urlprefix{URL }\fi
\providecommand{\bibinfo}[2]{#2}
\providecommand{\eprint}[2][]{\url{#2}}

\bibitem[{\citenamefont{Gy\"orffy et~al.}(1991)\citenamefont{Gy\"orffy, Stocks,
  Ginatempo, Johnson, Nicholson, Pinski, Staunton, Winter, Rafii-Tabar, Pendry
  et~al.}}]{Gyorffy1991a}
\bibinfo{author}{\bibfnamefont{B.~L.} \bibnamefont{Gy\"orffy}},
  \bibinfo{author}{\bibfnamefont{G.~M.} \bibnamefont{Stocks}},
  \bibinfo{author}{\bibfnamefont{B.}~\bibnamefont{Ginatempo}},
  \bibinfo{author}{\bibfnamefont{D.~D.} \bibnamefont{Johnson}},
  \bibinfo{author}{\bibfnamefont{D.~M.} \bibnamefont{Nicholson}},
  \bibinfo{author}{\bibfnamefont{F.~J.} \bibnamefont{Pinski}},
  \bibinfo{author}{\bibfnamefont{J.~B.} \bibnamefont{Staunton}},
  \bibinfo{author}{\bibfnamefont{H.}~\bibnamefont{Winter}},
  \bibinfo{author}{\bibfnamefont{H.}~\bibnamefont{Rafii-Tabar}},
  \bibinfo{author}{\bibfnamefont{J.~B.} \bibnamefont{Pendry}},
  \bibnamefont{et~al.}, \bibinfo{journal}{Philosophical Transactions: Physical
  Sciences and Engineering} \textbf{\bibinfo{volume}{334}}, \bibinfo{pages}{pp.
  515} (\bibinfo{year}{1991}), ISSN \bibinfo{issn}{09628428},
  \urlprefix\url{http://www.jstor.org/stable/53744}.

\bibitem[{\citenamefont{Kirkpatrick et~al.}(1970)\citenamefont{Kirkpatrick,
  Velick\'y, and Ehrenreich}}]{Kirkpatrick1970}
\bibinfo{author}{\bibfnamefont{S.}~\bibnamefont{Kirkpatrick}},
  \bibinfo{author}{\bibfnamefont{B.}~\bibnamefont{Velick\'y}},
  \bibnamefont{and}
  \bibinfo{author}{\bibfnamefont{H.}~\bibnamefont{Ehrenreich}},
  \bibinfo{journal}{Phys. Rev. B} \textbf{\bibinfo{volume}{1}},
  \bibinfo{pages}{3250} (\bibinfo{year}{1970}),
  \urlprefix\url{http://link.aps.org/doi/10.1103/PhysRevB.1.3250}.

\bibitem[{\citenamefont{Bansil}(1978)}]{Bansil1978}
\bibinfo{author}{\bibfnamefont{A.}~\bibnamefont{Bansil}},
  \bibinfo{journal}{Phys. Rev. Lett.} \textbf{\bibinfo{volume}{41}},
  \bibinfo{pages}{1670} (\bibinfo{year}{1978}),
  \urlprefix\url{http://link.aps.org/doi/10.1103/PhysRevLett.41.1670}.

\bibitem[{\citenamefont{Giuliano E~S and M}(1977)}]{Giuliano1978}
\bibinfo{author}{\bibfnamefont{G.~B.~L.} \bibnamefont{Giuliano E~S},
  \bibfnamefont{Ruggeri~R}} \bibnamefont{and}
  \bibinfo{author}{\bibfnamefont{S.~G.} \bibnamefont{M}}, in
  \emph{\bibinfo{booktitle}{Transition Metals}}, edited by
  \bibinfo{editor}{\bibfnamefont{J.~M.~P.} \bibnamefont{M~J G~Lee}}
  \bibnamefont{and} \bibinfo{editor}{\bibfnamefont{E.~F.} \bibnamefont{1978}}
  (\bibinfo{year}{1977}), vol.~\bibinfo{volume}{39} of
  \emph{\bibinfo{series}{Institute of Physics Conference Series}}.

\bibitem[{\citenamefont{Temmerman et~al.}(1978)\citenamefont{Temmerman,
  Gyorffy, and Stocks}}]{Temmerman1978}
\bibinfo{author}{\bibfnamefont{W.~M.} \bibnamefont{Temmerman}},
  \bibinfo{author}{\bibfnamefont{B.~L.} \bibnamefont{Gyorffy}},
  \bibnamefont{and} \bibinfo{author}{\bibfnamefont{G.~M.}
  \bibnamefont{Stocks}}, \bibinfo{journal}{Journal of Physics F: Metal Physics}
  \textbf{\bibinfo{volume}{8}}, \bibinfo{pages}{2461} (\bibinfo{year}{1978}),
  \urlprefix\url{http://stacks.iop.org/0305-4608/8/i=12/a=008}.

\bibitem[{\citenamefont{Matsubara}(1973)}]{Matsubara1973}
\bibinfo{author}{\bibfnamefont{T.}~\bibnamefont{Matsubara}},
  \bibinfo{journal}{Progress of Theoretical Physics Supplement}
  \textbf{\bibinfo{volume}{53}}, \bibinfo{pages}{202} (\bibinfo{year}{1973}),
  \urlprefix\url{http://ptp.ipap.jp/link?PTPS/53/202/}.

\bibitem[{\citenamefont{Staunton et~al.}(1984)\citenamefont{Staunton, Gyorffy,
  Pindor, Stocks, and Winter}}]{Staunton1984}
\bibinfo{author}{\bibfnamefont{J.}~\bibnamefont{Staunton}},
  \bibinfo{author}{\bibfnamefont{B.}~\bibnamefont{Gyorffy}},
  \bibinfo{author}{\bibfnamefont{A.}~\bibnamefont{Pindor}},
  \bibinfo{author}{\bibfnamefont{G.}~\bibnamefont{Stocks}}, \bibnamefont{and}
  \bibinfo{author}{\bibfnamefont{H.}~\bibnamefont{Winter}},
  \bibinfo{journal}{Journal of Magnetism and Magnetic Materials}
  \textbf{\bibinfo{volume}{45}}, \bibinfo{pages}{15 } (\bibinfo{year}{1984}),
  ISSN \bibinfo{issn}{0304-8853},
  \urlprefix\url{http://www.sciencedirect.com/science/article/pii/030488538490%
3676}.

\bibitem[{\citenamefont{Soven}(1967)}]{Soven1967}
\bibinfo{author}{\bibfnamefont{P.}~\bibnamefont{Soven}},
  \bibinfo{journal}{Phys. Rev.} \textbf{\bibinfo{volume}{156}},
  \bibinfo{pages}{809} (\bibinfo{year}{1967}).

\bibitem[{\citenamefont{Korringa}(1947)}]{Korringa1947}
\bibinfo{author}{\bibfnamefont{J.}~\bibnamefont{Korringa}},
  \bibinfo{journal}{Physica} \textbf{\bibinfo{volume}{13}}, \bibinfo{pages}{392
  } (\bibinfo{year}{1947}), ISSN \bibinfo{issn}{0031-8914},
  \urlprefix\url{http://www.sciencedirect.com/science/article/pii/003189144790%
013X}.

\bibitem[{\citenamefont{Kohn and Rostoker}(1954)}]{Kohn1954}
\bibinfo{author}{\bibfnamefont{W.}~\bibnamefont{Kohn}} \bibnamefont{and}
  \bibinfo{author}{\bibfnamefont{N.}~\bibnamefont{Rostoker}},
  \bibinfo{journal}{Phys. Rev.} \textbf{\bibinfo{volume}{94}},
  \bibinfo{pages}{1111} (\bibinfo{year}{1954}).

\bibitem[{\citenamefont{Gy\"orffy}(1978)}]{Gyorffy1978}
\bibinfo{author}{\bibfnamefont{B.}~\bibnamefont{Gy\"orffy}},
  \emph{\bibinfo{title}{Electrons in disordered metals and at metallic
  surfaces}} (\bibinfo{publisher}{Plenum Press}, \bibinfo{year}{1978}).

\bibitem[{\citenamefont{Temmerman and Szotek}(1987)}]{Temmerman1987}
\bibinfo{author}{\bibfnamefont{W.~M.} \bibnamefont{Temmerman}}
  \bibnamefont{and} \bibinfo{author}{\bibfnamefont{Z.}~\bibnamefont{Szotek}},
  \bibinfo{journal}{Computer Physics reports} \textbf{\bibinfo{volume}{5}},
  \bibinfo{pages}{173} (\bibinfo{year}{1987}), ISSN \bibinfo{issn}{0167-7977},
  \urlprefix\url{http://www.sciencedirect.com/science/article/pii/016779778790%
0062}.

\bibitem[{\citenamefont{Mills and Ratanavararaksa}(1978)}]{Mills1978}
\bibinfo{author}{\bibfnamefont{R.}~\bibnamefont{Mills}} \bibnamefont{and}
  \bibinfo{author}{\bibfnamefont{P.}~\bibnamefont{Ratanavararaksa}},
  \bibinfo{journal}{Phys. Rev. B} \textbf{\bibinfo{volume}{18}},
  \bibinfo{pages}{5291} (\bibinfo{year}{1978}).

\bibitem[{\citenamefont{Mills et~al.}(1983)\citenamefont{Mills, Gray, and
  Kaplan}}]{Mills1983}
\bibinfo{author}{\bibfnamefont{R.}~\bibnamefont{Mills}},
  \bibinfo{author}{\bibfnamefont{L.~J.} \bibnamefont{Gray}}, \bibnamefont{and}
  \bibinfo{author}{\bibfnamefont{T.}~\bibnamefont{Kaplan}},
  \bibinfo{journal}{Phys. Rev. B} \textbf{\bibinfo{volume}{27}},
  \bibinfo{pages}{3252} (\bibinfo{year}{1983}).

\bibitem[{\citenamefont{Velick\'y et~al.}(1968)\citenamefont{Velick\'y,
  Kirkpatrick, and Ehrenreich}}]{Velicky1968}
\bibinfo{author}{\bibfnamefont{B.}~\bibnamefont{Velick\'y}},
  \bibinfo{author}{\bibfnamefont{S.}~\bibnamefont{Kirkpatrick}},
  \bibnamefont{and}
  \bibinfo{author}{\bibfnamefont{H.}~\bibnamefont{Ehrenreich}},
  \bibinfo{journal}{Phys. Rev.} \textbf{\bibinfo{volume}{175}},
  \bibinfo{pages}{747} (\bibinfo{year}{1968}),
  \urlprefix\url{http://link.aps.org/doi/10.1103/PhysRev.175.747}.

\bibitem[{\citenamefont{Levin et~al.}(1970)\citenamefont{Levin, Velick\'y, and
  Ehrenreich}}]{Levin1970}
\bibinfo{author}{\bibfnamefont{K.}~\bibnamefont{Levin}},
  \bibinfo{author}{\bibfnamefont{B.}~\bibnamefont{Velick\'y}},
  \bibnamefont{and}
  \bibinfo{author}{\bibfnamefont{H.}~\bibnamefont{Ehrenreich}},
  \bibinfo{journal}{Phys. Rev. B} \textbf{\bibinfo{volume}{2}},
  \bibinfo{pages}{1771} (\bibinfo{year}{1970}),
  \urlprefix\url{http://link.aps.org/doi/10.1103/PhysRevB.2.1771}.

\bibitem[{\citenamefont{Čápek}(1971)}]{Capek1971}
\bibinfo{author}{\bibfnamefont{V.}~\bibnamefont{Čápek}},
  \bibinfo{journal}{physica status solidi (b)} \textbf{\bibinfo{volume}{43}},
  \bibinfo{pages}{61} (\bibinfo{year}{1971}), ISSN \bibinfo{issn}{1521-3951},
  \urlprefix\url{http://dx.doi.org/10.1002/pssb.2220430106}.

\bibitem[{\citenamefont{Rowlands}(2009)}]{Rowlands2009}
\bibinfo{author}{\bibfnamefont{D.~A.} \bibnamefont{Rowlands}},
  \bibinfo{journal}{Reports on Progress in Physics}
  \textbf{\bibinfo{volume}{72}}, \bibinfo{pages}{086501}
  (\bibinfo{year}{2009}),
  \urlprefix\url{http://stacks.iop.org/0034-4885/72/i=8/a=086501}.

\bibitem[{\citenamefont{Pindor et~al.}(1983)\citenamefont{Pindor, Temmerman,
  and Gy\"{o}rffy}}]{Pindor1983}
\bibinfo{author}{\bibfnamefont{A.~J.} \bibnamefont{Pindor}},
  \bibinfo{author}{\bibfnamefont{W.~M.} \bibnamefont{Temmerman}},
  \bibnamefont{and} \bibinfo{author}{\bibfnamefont{B.~L.}
  \bibnamefont{Gy\"{o}rffy}}, \bibinfo{journal}{Journal of Physics F: Metal
  Physics} \textbf{\bibinfo{volume}{13}}, \bibinfo{pages}{1627}
  (\bibinfo{year}{1983}),
  \urlprefix\url{http://stacks.iop.org/0305-4608/13/i=8/a=009}.

\bibitem[{\citenamefont{Jarrell and Krishnamurthy}(2001)}]{Jarrell2001}
\bibinfo{author}{\bibfnamefont{M.}~\bibnamefont{Jarrell}} \bibnamefont{and}
  \bibinfo{author}{\bibfnamefont{H.~R.} \bibnamefont{Krishnamurthy}},
  \bibinfo{journal}{Phys. Rev. B} \textbf{\bibinfo{volume}{63}},
  \bibinfo{pages}{125102} (\bibinfo{year}{2001}).

\bibitem[{\citenamefont{Rowlands et~al.}(2003)\citenamefont{Rowlands, Staunton,
  and Gy\"orffy}}]{Rowlands2003}
\bibinfo{author}{\bibfnamefont{D.~A.} \bibnamefont{Rowlands}},
  \bibinfo{author}{\bibfnamefont{J.~B.} \bibnamefont{Staunton}},
  \bibnamefont{and} \bibinfo{author}{\bibfnamefont{B.~L.}
  \bibnamefont{Gy\"orffy}}, \bibinfo{journal}{Phys. Rev. B}
  \textbf{\bibinfo{volume}{67}}, \bibinfo{pages}{115109}
  (\bibinfo{year}{2003}).

\bibitem[{\citenamefont{Rowlands}(2004)}]{Rowlands2004}
\bibinfo{author}{\bibfnamefont{D.}~\bibnamefont{Rowlands}}, Ph.D. thesis,
  \bibinfo{school}{University of Warwick} (\bibinfo{year}{2004}).

\bibitem[{\citenamefont{Rowlands}(2006)}]{Rowlands2006}
\bibinfo{author}{\bibfnamefont{D.}~\bibnamefont{Rowlands}},
  \bibinfo{journal}{Psi-k newsletter} \textbf{\bibinfo{volume}{77}}
  (\bibinfo{year}{2006}).

\bibitem[{\citenamefont{Rowlands et~al.}(2008)\citenamefont{Rowlands, Zhang,
  and Gonis}}]{Rowlands2008}
\bibinfo{author}{\bibfnamefont{D.~A.} \bibnamefont{Rowlands}},
  \bibinfo{author}{\bibfnamefont{X.-G.} \bibnamefont{Zhang}}, \bibnamefont{and}
  \bibinfo{author}{\bibfnamefont{A.}~\bibnamefont{Gonis}},
  \bibinfo{journal}{Phys. Rev. B} \textbf{\bibinfo{volume}{78}},
  \bibinfo{pages}{115119} (\bibinfo{year}{2008}).

\bibitem[{\citenamefont{Rowlands et~al.}(2006)\citenamefont{Rowlands, Ernst,
  Gy\"orffy, and Staunton}}]{Rowlands2006a}
\bibinfo{author}{\bibfnamefont{D.~A.} \bibnamefont{Rowlands}},
  \bibinfo{author}{\bibfnamefont{A.}~\bibnamefont{Ernst}},
  \bibinfo{author}{\bibfnamefont{B.~L.} \bibnamefont{Gy\"orffy}},
  \bibnamefont{and} \bibinfo{author}{\bibfnamefont{J.~B.}
  \bibnamefont{Staunton}}, \bibinfo{journal}{Phys. Rev. B}
  \textbf{\bibinfo{volume}{73}}, \bibinfo{pages}{165122}
  (\bibinfo{year}{2006}),
  \urlprefix\url{http://link.aps.org/doi/10.1103/PhysRevB.73.165122}.

\bibitem[{\citenamefont{Tulip et~al.}(2006)\citenamefont{Tulip, Staunton,
  Rowlands, Gy\"orffy, Bruno, and Ginatempo}}]{Tulip2006}
\bibinfo{author}{\bibfnamefont{P.~R.} \bibnamefont{Tulip}},
  \bibinfo{author}{\bibfnamefont{J.~B.} \bibnamefont{Staunton}},
  \bibinfo{author}{\bibfnamefont{D.~A.} \bibnamefont{Rowlands}},
  \bibinfo{author}{\bibfnamefont{B.~L.} \bibnamefont{Gy\"orffy}},
  \bibinfo{author}{\bibfnamefont{E.}~\bibnamefont{Bruno}}, \bibnamefont{and}
  \bibinfo{author}{\bibfnamefont{B.}~\bibnamefont{Ginatempo}},
  \bibinfo{journal}{Phys. Rev. B} \textbf{\bibinfo{volume}{73}},
  \bibinfo{pages}{205109} (\bibinfo{year}{2006}).

\bibitem[{\citenamefont{K\"odderitzsch
  et~al.}(2007)\citenamefont{K\"odderitzsch, Ebert, Rowlands, and
  Ernst}}]{Koedderitzsch2007}
\bibinfo{author}{\bibfnamefont{D.}~\bibnamefont{K\"odderitzsch}},
  \bibinfo{author}{\bibfnamefont{H.}~\bibnamefont{Ebert}},
  \bibinfo{author}{\bibfnamefont{D.~A.} \bibnamefont{Rowlands}},
  \bibnamefont{and} \bibinfo{author}{\bibfnamefont{A.}~\bibnamefont{Ernst}},
  \bibinfo{journal}{New Journal of Physics} \textbf{\bibinfo{volume}{9}},
  \bibinfo{pages}{81} (\bibinfo{year}{2007}),
  \urlprefix\url{http://stacks.iop.org/1367-2630/9/i=4/a=081}.

\bibitem[{\citenamefont{Marmodoro and Staunton}(2011)}]{Marmodoro2011}
\bibinfo{author}{\bibfnamefont{A.}~\bibnamefont{Marmodoro}} \bibnamefont{and}
  \bibinfo{author}{\bibfnamefont{J.~B.} \bibnamefont{Staunton}},
  \bibinfo{journal}{Journal of Physics: Conference Series}
  (\bibinfo{year}{2011}),
  \urlprefix\url{http://stacks.iop.org/1742-6596/286/i=1/a=012033}.

\bibitem[{\citenamefont{Khalaff and Schubert}(1974)}]{Khalaff1974}
\bibinfo{author}{\bibfnamefont{K.}~\bibnamefont{Khalaff}} \bibnamefont{and}
  \bibinfo{author}{\bibfnamefont{K.}~\bibnamefont{Schubert}},
  \bibinfo{journal}{Journal of the Less Common Metals}
  \textbf{\bibinfo{volume}{35}}, \bibinfo{pages}{341 } (\bibinfo{year}{1974}),
  ISSN \bibinfo{issn}{0022-5088},
  \urlprefix\url{http://www.sciencedirect.com/science/article/pii/002250887490%
2471}.

\bibitem[{\citenamefont{{Anand} et~al.}(2003)\citenamefont{{Anand}, {Taylor},
  {Nazarov}, {Shu}, {Mao}, and {Hemley}}}]{Anand2003}
\bibinfo{author}{\bibfnamefont{M.}~\bibnamefont{{Anand}}},
  \bibinfo{author}{\bibfnamefont{L.~A.} \bibnamefont{{Taylor}}},
  \bibinfo{author}{\bibfnamefont{M.~A.} \bibnamefont{{Nazarov}}},
  \bibinfo{author}{\bibfnamefont{J.}~\bibnamefont{{Shu}}},
  \bibinfo{author}{\bibfnamefont{H.-K.} \bibnamefont{{Mao}}}, \bibnamefont{and}
  \bibinfo{author}{\bibfnamefont{R.~J.} \bibnamefont{{Hemley}}}, in
  \emph{\bibinfo{booktitle}{Lunar and Planetary Institute Science Conference
  Abstracts}}, edited by \bibinfo{editor}{\bibnamefont{{S.~Mackwell \&
  E.~Stansbery}}} (\bibinfo{year}{2003}), vol.~\bibinfo{volume}{34} of
  \emph{\bibinfo{series}{Lunar and Planetary Institute Science Conference
  Abstracts}}, p. \bibinfo{pages}{1818}.

\bibitem[{\citenamefont{{Hiltl} et~al.}(2011)\citenamefont{{Hiltl}, {Bauer},
  {Ernstson}, {Mayer}, {Neumair}, and {Rappengl{\"u}ck}}}]{Hiltl2011}
\bibinfo{author}{\bibfnamefont{M.}~\bibnamefont{{Hiltl}}},
  \bibinfo{author}{\bibfnamefont{F.}~\bibnamefont{{Bauer}}},
  \bibinfo{author}{\bibfnamefont{K.}~\bibnamefont{{Ernstson}}},
  \bibinfo{author}{\bibfnamefont{W.}~\bibnamefont{{Mayer}}},
  \bibinfo{author}{\bibfnamefont{A.}~\bibnamefont{{Neumair}}},
  \bibnamefont{and} \bibinfo{author}{\bibfnamefont{M.~A.}
  \bibnamefont{{Rappengl{\"u}ck}}}, in \emph{\bibinfo{booktitle}{Lunar and
  Planetary Institute Science Conference Abstracts}} (\bibinfo{year}{2011}),
  vol.~\bibinfo{volume}{42} of \emph{\bibinfo{series}{Lunar and Planetary
  Institute Science Conference Abstracts}}, p. \bibinfo{pages}{1391}.

\bibitem[{\citenamefont{Johnson et~al.}(1986)\citenamefont{Johnson, Nicholson,
  Pinski, Gyorffy, and Stocks}}]{Johnson1986}
\bibinfo{author}{\bibfnamefont{D.~D.} \bibnamefont{Johnson}},
  \bibinfo{author}{\bibfnamefont{D.~M.} \bibnamefont{Nicholson}},
  \bibinfo{author}{\bibfnamefont{F.~J.} \bibnamefont{Pinski}},
  \bibinfo{author}{\bibfnamefont{B.~L.} \bibnamefont{Gyorffy}},
  \bibnamefont{and} \bibinfo{author}{\bibfnamefont{G.~M.}
  \bibnamefont{Stocks}}, \bibinfo{journal}{Phys. Rev. Lett.}
  \textbf{\bibinfo{volume}{56}}, \bibinfo{pages}{2088} (\bibinfo{year}{1986}).

\bibitem[{\citenamefont{Faulkner and Stocks}(1980)}]{Faulkner1980}
\bibinfo{author}{\bibfnamefont{J.~S.} \bibnamefont{Faulkner}} \bibnamefont{and}
  \bibinfo{author}{\bibfnamefont{G.~M.} \bibnamefont{Stocks}},
  \bibinfo{journal}{Phys. Rev. B} \textbf{\bibinfo{volume}{21}},
  \bibinfo{pages}{3222} (\bibinfo{year}{1980}),
  \urlprefix\url{http://10.1103/PhysRevB.21.3222}.

\bibitem[{\citenamefont{Gonis}(1992)}]{Gonis1992}
\bibinfo{author}{\bibfnamefont{A.}~\bibnamefont{Gonis}},
  \emph{\bibinfo{title}{Green Functions for Ordered and Disordered Systems
  (Studies in Mathematical Physics)}} (\bibinfo{publisher}{North-Holland},
  \bibinfo{year}{1992}), ISBN \bibinfo{isbn}{0444889868},
  \urlprefix\url{http://researchbooks.org/0444889868}.

\bibitem[{\citenamefont{Ham and Segall}(1961)}]{Ham1961}
\bibinfo{author}{\bibfnamefont{F.~S.} \bibnamefont{Ham}} \bibnamefont{and}
  \bibinfo{author}{\bibfnamefont{B.}~\bibnamefont{Segall}},
  \bibinfo{journal}{Phys. Rev.} \textbf{\bibinfo{volume}{124}},
  \bibinfo{pages}{1786} (\bibinfo{year}{1961}).

\bibitem[{\citenamefont{Kubo}(1962)}]{Kubo1962}
\bibinfo{author}{\bibfnamefont{R.}~\bibnamefont{Kubo}},
  \bibinfo{journal}{Journal of the Physical Society of Japan}
  \textbf{\bibinfo{volume}{17}}, \bibinfo{pages}{1100} (\bibinfo{year}{1962}),
  \urlprefix\url{http://jpsj.ipap.jp/link?JPSJ/17/1100/}.

\bibitem[{\citenamefont{Ziman}(1979)}]{Ziman1979}
\bibinfo{author}{\bibfnamefont{J.~M.} \bibnamefont{Ziman}},
  \emph{\bibinfo{title}{Models of disorder: the theoretical physics of
  homogeneously disordered systems}} (\bibinfo{publisher}{Cambridge University
  Press}, \bibinfo{year}{1979}).

\end{thebibliography}
\end{document}